\documentclass[preprint,10pt]{elsarticle}
\pdfoutput=1


\makeatletter
\def\ps@pprintTitle{%
   \let\@oddhead\@empty
   \let\@evenhead\@empty
   \def\@oddfoot{\reset@font\hfil\thepage\hfil}
   \let\@evenfoot\@oddfoot
}
\makeatother

\usepackage{titlesec}
\usepackage{xcolor}
\usepackage{enumitem}
\usepackage{hyperref}
\hypersetup{colorlinks=true,urlcolor=black}
\usepackage[T1]{fontenc}\usepackage{lmodern}

\usepackage{mathtools,amsfonts,amssymb}
\usepackage{caption}
\usepackage{subcaption}
\usepackage{booktabs}
\usepackage{pifont}
\usepackage{placeins}

\definecolor{green}{rgb}{0,0.5,0}

\usepackage{tikz}
\usetikzlibrary{calc}
\usepackage{tkz-euclide}
\usepackage{calc}
\usepackage{tikz,pgfplots}
\pgfplotsset{compat=1.12}
\usetikzlibrary{arrows.meta,3d,matrix,positioning,decorations.markings}

\newcommand{\cmark}{\ding{51}}%
\newcommand{\xmark}{\ding{55}}

\newcommand{\pz}{\phantom{0}}
\newcommand{\ntriangles}{n_t}
\newcommand{\rank}{m_b}
\newcommand{\nbasis}{n_b}
\newcommand{\srcidx}{j}
\newcommand{\testidx}{i}

\makeatletter
\tikzoption{canvas is xy plane at z}[]{%
  \def\tikz@plane@origin{\pgfpointxyz{0}{0}{#1}}%
  \def\tikz@plane@x{\pgfpointxyz{1}{0}{#1}}%
  \def\tikz@plane@y{\pgfpointxyz{0}{1}{#1}}%
  \tikz@canvas@is@plane
}
\makeatother

\makeatletter
\tikzoption{canvas is plane}[]{\@setOxy#1}
\def\@setOxy O(#1,#2,#3)x(#4,#5,#6)y(#7,#8,#9)%
  {\def\tikz@plane@origin{\pgfpointxyz{#1}{#2}{#3}}%
   \def\tikz@plane@x{\pgfpointxyz{#4}{#5}{#6}}%
   \def\tikz@plane@y{\pgfpointxyz{#7}{#8}{#9}}%
   \tikz@canvas@is@plane
  }
\makeatother 

\definecolor{orange}{rgb}{1,0.5,0}
\definecolor{green}{rgb}{0,0.5,0}
\definecolor{purple}{rgb}{0.5,0,0.5}
\newcommand{\reviewerOne}[1]{#1}
\newcommand{\rereading}[1]{#1}
\newcommand{\reviewerTwo}[1]{#1}
\usepackage[margin=1in]{geometry}
\biboptions{sort&compress}
\begin{document}

\begin{frontmatter}
\title{Manufactured Solutions for the Method-of-Moments Implementation of the Electric-Field Integral Equation}

\author[freno]{Brian A.\ Freno}
\ead{bafreno@sandia.gov}
\author[freno]{Neil R.\ Matula}
\author[freno]{\rereading{William A.\ Johnson}}

\address[freno]{Sandia National Laboratories, Albuquerque, NM 87185}

\begin{abstract}
Though the method-of-moments implementation of the electric-field integral equation plays an important role in computational electromagnetics, it provides many code-verification challenges due to the different sources of numerical error.  In this paper, we provide an approach through which we can apply the method of manufactured solutions to isolate and verify the solution-discretization error.  We accomplish this by manufacturing both the surface current and the Green's function.  Because the arising equations are poorly conditioned, we reformulate them as a set of constraints for an optimization problem that selects the solution closest to the manufactured solution.  We demonstrate the effectiveness of this approach for cases with and without coding errors.
\end{abstract}


\end{frontmatter}

\section{Introduction}

The method-of-moments (MoM) implementation of the electric-field integral equation (EFIE) is a useful technique for numerically modeling electromagnetic scattering and radiation problems.  Through this approach, the surface of the electromagnetic scatterer is discretized using planar or curvilinear mesh elements, and four-dimensional integrals are evaluated over two-dimensional source and test elements.  However, the presence of a Green's function in these equations yields scalar- and vector-potential terms with singularities when the test and source elements share one or more edges or vertices and near-singularities when they are otherwise close.  Many approaches have been developed to address the singularity and near-singularity for the inner, source-element integral~\cite{graglia_1993,wilton_1984,rao_1982,khayat_2005,fink_2008,khayat_2008,vipiana_2011,vipiana_2012,botha_2013,rivero_2019}, as well as for the outer, test-element integral~\cite{vipiana_2013,polimeridis_2013,wilton_2017,rivero_2019b,freno_em}.

For computational physics codes in general, it is necessary to assess the implementation and the suitability of the underlying models in order to develop confidence in the simulation results.  These assessments typically fall into two complementary categories: verification and validation.  Validation evaluates the appropriateness of the models instantiated in the code for representing the relevant physical phenomena, and is typically performed through comparison with experimental data.  Verification, on the other hand, assesses the correctness of the numerical solutions produced by the code, through comparison with the expected theoretical behavior of the implemented numerical methods.  Following Roache~\cite{roache_1998},  Salari and Knupp~\cite{salari_2000}, and Oberkampf and Roy~\cite{oberkampf_2010}, verification can be further divided into the activities of code verification and solution verification.  Solution verification involves the estimation of the numerical error for a particular simulation, whereas code verification assesses the correctness of the implementation of the numerical methods within the code.  A review of code and solution verification is presented by Roy~\cite{roy_2005}.

Code verification is the focus of this paper.  In general, codes that approximately solve systems of differential, integral, or integro-differential equations can only be verified by using them to solve problems with known solutions~\cite{roache_2001}.  The discretization of the governing equations necessarily incurs some truncation error, and thus the approximate solutions produced from the discretized equations will incur an associated discretization error.  If the solution to the problem is known, a measure of the discretization error (typically a discrete norm thereof) may be evaluated directly from the approximate solution.  In the most basic sense of verification, if the discretization error tends to zero as the discretization is refined, the consistency of the code is verified~\cite{roache_1998}.  This may be taken a step further by examining not only consistency, but the rate at which the error decreases as the discretization is refined, thereby verifying the order of accuracy of the discretization scheme.  The correctness of the numerical-method implementation may then be verified by comparing the expected and observed orders of accuracy obtained from numerous test cases with known solutions.

Exact solutions to systems of engineering interest are rare, and those that do exist often require dramatic simplifications to both the domain geometry and the equations themselves in order to obtain a tractable problem.  Hence, manufactured solutions are frequently employed to produce problems of sufficient complexity with known solutions~\cite{roache_2001}.  The method of manufactured solutions (MMS) is a general technique for constructing problems of arbitrary complexity with known solutions.  One begins this process in reverse by manufacturing the desired solution.  In principle, this manufactured solution (MS) may be any function, but several properties are desirable~\cite{salari_2000}:

\begin{enumerate}[leftmargin=*,labelindent=\parindent]
\item \label{prop:elementary} The MS should consist of combinations of elementary functions, such as polynomial, trigonometric, and exponential functions.  This not only simplifies derivations and implementation, but ensures that the MS (and its derivatives) will be representable to sufficient precision within the tested code.

\item The MS should be sufficiently smooth, such that the error incurred by the discretization is small on relatively coarse meshes.  This ensures that the order of accuracy may be estimated with minimal computational expense.

\item The MS should be general enough that all terms of the governing equations are exercised.

\item The MS should have a sufficient number of nontrivial derivatives, such that the expected order of accuracy of the discretization can be observed.  In the most ideal case, the solution will have an infinite number of nontrivial derivatives.

\item Since the robustness of the code is not the primary concern, the MS should not have any features that inhibit the solution of the discretized equations.
\end{enumerate}

Once a solution is manufactured, it is substituted directly into the governing equations.  
In general, the MS is not expected to satisfy the governing equations.  Instead, a residual term appears, which quantifies the deviation from the satisfaction of the equations.  If this residual is added to the governing equations as a source term, the resulting equations will be exactly satisfied by the MS.  Concerns are immediately raised regarding uniqueness of the solution to the manufactured problem, but this has rarely been found to cause difficulties in practice~\cite{roache_2001}.  The result of this process is a problem, of arbitrary complexity, for which an exact solution is known.  

The code to be verified is then modified to support the additional source term and may then be verified by comparing the simulation result for the manufactured problem against the known solution.  
\reviewerOne{Ideally, the source term is computed analytically; however, when this is not possible, the source term needs to be computed consistently and at least as accurately as the numerical methods being verified.  Otherwise, the error in the source term will overshadow that of the numerical methods, contaminating the verification assessment.}

MMS is purely a mathematical process; the physics of the problem need not be considered, \reviewerTwo{provided the MS remains within the bounds of validity for the underlying algorithms}.  This enables the user to avoid difficulties that would normally complicate the solution.

Code verification has been performed on computational physics codes associated with several physics disciplines, including fluid dynamics~\cite{roy_2004,bond_2007,veluri_2010,oliver_2012,eca_2016,freno_2021}, solid mechanics~\cite{chamberland_2010}, fluid--structure interaction~\cite{etienne_2012}, heat transfer in fluid--solid interaction~\cite{veeraragavan_2016}, multiphase flows~\cite{brady_2012}, radiation hydrodynamics~\cite{mcclarren_2008}, electrodynamics~\cite{ellis_2009}, and ablation~\cite{amar_2008,amar_2009,amar_2011,freno_ablation}.  
However, existing literature contains few instances of MMS being used in the verification of software for integral equations.  This is due to the simple fact that, while analytical differentiation is a straightforward exercise, analytical integration is not always possible.  Hence, the residual source term arising from the manufactured solution may not be representable in closed form, and its implementation may be accompanied by numerical techniques that carry their own uncertainties.  Furthermore, in many applications, such as the MoM implementation of the EFIE, singular integrals appear, which can further complicate the numerical evaluation of the source term.  Therefore, much of the elegance, simplicity, and instilled confidence of MMS is lost when applied to integral equations in a straightforward manner, and, as a result, effective implementation of MMS in the context of boundary element codes is an open subject of research.  

The most substantial effort thus far for integral equations in computational electromagnetics has been the work of Marchand et al.~\cite{marchand_2013,marchand_2014}, in which the authors calculate the MMS source terms for the EFIE using numerical techniques.  While the quadrature error can be driven low enough that \reviewerTwo{mesh-refinement} studies become feasible, the presence of this additional error often places a lower bound on the discretization error that can be obtained, and therefore limits the scope of the \reviewerTwo{mesh-refinement} study, in addition to being undesirable for the aforementioned reasons.  To further complicate matters, for the MoM implementation of the EFIE, the discretized equations can easily become ill-conditioned~\cite{adrian_2019}.

For the MoM implementation of the EFIE, numerical error is introduced from three sources:
\begin{enumerate}[leftmargin=*,labelindent=\parindent]
\item \label{err:dom_disc} \textbf{Domain discretization.} While planar surfaces can be represented exactly by planar elements, the approximation of curved surfaces with planar elements introduces a second-order numerical error~\cite[Chap.~3]{warnick_2008}.  This error can be reduced by employing curved elements, but we restrict the scope of this work to planar elements.

\item \label{err:sol_disc} \textbf{Solution discretization.} Common in the solution to differential, integral, and integro-differential equations, the approximation of the solution in terms of a finite number of basis functions, or alternatively the approximation of the underlying equation operators in terms of a finite amount of solution queries, is the most common contributor to the numerical error. 

\item \label{err:num_int}\textbf{Numerical integration.} 
As previously stated, the analytical evaluation of the integrals in integral equations is not always possible.  For well-behaved integrals, quadrature rules or other integration methods can be used, with the expectation that the associated numerical error is at least of the same order as that arising from Error Source~\ref{err:sol_disc}.  A less rigorous expectation is that the error from numerical integration decreases as the fidelity of the numerical integration algorithm is increased (e.g., increasing the number of quadrature points).  However, for singular or nearly singular integrals, such convergence is not assured~\cite{freno_quad}.

\end{enumerate}
It is important to account for these error sources when performing code verification.  Table~\ref{tab:combinations} lists the possible combinations of these error sources.

\begin{table}[!b]
\centering
\begin{tabular}{c c c c c }
\toprule
& \multicolumn{3}{c}{Error Source} \\
 \cmidrule(lr){2-4}
Combination & 
\hspace{0.5em}\ref{err:dom_disc}\hspace{0.5em} & 
\hspace{0.5em}\ref{err:sol_disc}\hspace{0.5em} & 
\hspace{0.5em}\ref{err:num_int} \hspace{0.5em} &
Verifiable?\\
\midrule
1 & \cmark &        &        & \color{red}\xmark \\
2 &        & \cmark &        & \color{green}\cmark \\
3 &        &        & \cmark & \color{red}\xmark \\
4 & \cmark & \cmark &        & \color{green}\cmark \\
5 &        & \cmark & \cmark & \color{green}\cmark \\
6 & \cmark &        & \cmark & \color{red}\xmark \\
7 & \cmark & \cmark & \cmark & \color{green}\cmark \\
\bottomrule
\end{tabular}
\caption{Possible combinations of error sources.}
\vskip-\dp\strutbox
\label{tab:combinations}
\end{table}

Error Source~\ref{err:dom_disc} can be eliminated for planar surfaces, and Error Source~\ref{err:num_int} can be eliminated if the numerical integration can be performed exactly.  However, for certain solution discretization choices, such as the use of the Rao--Wilton--Glisson basis functions, nontrivial solutions cannot be exactly represented across different meshes.  
Therefore, we cannot verify Combinations 1, 3, or 6.  
Exact solutions exist for curved surfaces, such as the Mie solution to Maxwell's equations for a sphere, permitting Combination 7 to be verified.
Combination 5 can also be verified; however, while Error Source~\ref{err:num_int} can be easily studied for nonsingular integrands, for singular and nearly singular integrands, Error Source~\ref{err:num_int} is most likely better studied using unit tests, due to the computationally expensive adaptive integration techniques required to compute accurate reference solutions~\cite{freno_quad,freno_em}.

In order for Combinations 2 and 4 to be verified, the numerical integration needs to be exact, which is not possible for the Green's function with nontrivial solutions.  Therefore, the purpose of this paper is to present a code-verification technique for verifying Combination 2 in the MoM implementation of the EFIE.  

Through this approach, we manufacture the surface current and Green's function and only consider surfaces that can be represented by planes. The planar surfaces avoid Error Source~\ref{err:dom_disc}, and, by manufacturing the two functions, we can quickly and exactly compute the integrals, avoiding Error Source~\ref{err:num_int}.  This enables us to isolate Error Source~\ref{err:sol_disc}.  Using the approach in this paper, one could verify Combination 4 as well.

Despite permitting analytical evaluation of the integrals, the manufactured Green's function further worsens the conditioning of the discretized equations, permitting multiple solutions.  To mitigate this challenge, we reduce the equations to a set of constraints and admit the solution closest to the MS that satisfies these constraints.

This paper is organized as follows.  In Section~\ref{sec:efie}, we describe the MoM implementation of the EFIE.  In Section~\ref{sec:mms}, we describe the challenges of using MMS with the MoM implementation of the EFIE, and we describe our approach to mitigating them.  In Section~\ref{sec:results}, we demonstrate the effectiveness of our approach for cases with and without coding errors.  In Section~\ref{sec:conclusions}, we summarize this work and provide an outlook for future work.

\section{The Method-of-Moments Implementation of the EFIE}
\label{sec:efie}

In time-harmonic form, the scattered electric field $\mathbf{E}^\mathcal{S}$ can be computed from the surface current by
\begin{align}\rereading{
\mathbf{E}^\mathcal{S} = -\left(j\omega\mathbf{A}+\nabla\Phi\right)}, 
\label{eq:Es}
\end{align}
where the magnetic vector potential $\mathbf{A}$ is defined by
\begin{align}
\mathbf{A}(\mathbf{x})= \mu \int_{S'} \mathbf{J} (\mathbf{x}')G(\mathbf{x},\mathbf{x}')dS',
\label{eq:A}
\end{align}
and, \rereading{by employing the Lorenz gauge condition and the continuity equation}, the electric scalar potential $\Phi$ is defined by 
\begin{align}
\Phi(\mathbf{x})=  \frac{j}{\epsilon\omega} \int_{S'} \nabla'\cdot\mathbf{J}(\mathbf{x}')G(\mathbf{x},\mathbf{x}')dS'.
\label{eq:Phi}
\end{align}
In~\eqref{eq:A} and~\eqref{eq:Phi}, \reviewerOne{the integration domain is the surface $S$ of a perfectly conducting scatterer.  Additionally,} $\mathbf{J}$ is the surface current, $\mu$ and $\epsilon$ are the permeability and permittivity of surrounding medium, and $G$ is the Green's function 
\begin{align}
G(\mathbf{x},\mathbf{x}') = \frac{e^{-jkR}}{4\pi R},
\label{eq:G}
\end{align}
where $R=|\mathbf{x}-\mathbf{x}'|$, and $k=\omega\sqrt{\mu\epsilon}$ is the wave number.  If $S$ is open, the component of $\mathbf{J}$ normal to the boundary of $S$ must vanish on the boundary of $S$ \rereading{to reflect that the total current is zero}.

The total electric field $\mathbf{E}$ is the sum of the incident electric field $\mathbf{E}^\mathcal{I}$ \rereading{(which induces $\mathbf{J}$)} and $\mathbf{E}^\mathcal{S}$.  On $S$, the tangential component of $\mathbf{E}$ is zero, such that
\begin{align}\reviewerOne{
\mathbf{E}_t^\mathcal{S}=-\mathbf{E}_t^\mathcal{I}},
\label{eq:tan_BC}
\end{align}
\reviewerOne{%
where the subscript $t$ denotes the tangential component.
Substituting~\eqref{eq:Es} into~\eqref{eq:tan_BC}}, we can compute $\mathbf{J}$ from $\mathbf{E}^\mathcal{I}$:
\begin{align}
\mathbf{E}_t^\mathcal{I} =\left(j\omega\mathbf{A} + \nabla\Phi\right)_t. 
\label{eq:tan}
\end{align}

To solve~\eqref{eq:tan} \reviewerOne{for $\mathbf{J}$, we discretize $S$ with a mesh composed of triangular elements and approximate $\mathbf{J}$ with $\mathbf{J}_h$ in terms of the Rao--Wilton--Glisson (RWG) basis functions $\boldsymbol{\Lambda}_{\srcidx}(\mathbf{x})$~\cite{rao_1982}:
\begin{align}
\mathbf{J}_h(\mathbf{x}) = \sum_{\srcidx=1}^{\nbasis} J_{\srcidx} \boldsymbol{\Lambda}_{\srcidx}(\mathbf{x}),
\label{eq:J_approx}
\end{align}}%
where $\nbasis$ is the \rereading{total number of basis functions. 
The RWG basis functions are second-order accurate~\cite[pp.\ 155--156]{warnick_2008}, and are defined for a triangle pair} by
%
%
\begin{align*}
\boldsymbol{\Lambda}_{\srcidx}(\mathbf{x}) = \left\{
\begin{matrix}
\displaystyle\frac{\ell_{\srcidx}}{2A_{\srcidx}^+}\boldsymbol{\rho}_{\srcidx}^+, & \text{for }\mathbf{x}\in T_{\srcidx}^+ \\[1em]
\displaystyle\frac{\ell_{\srcidx}}{2A_{\srcidx}^-}\boldsymbol{\rho}_{\srcidx}^-, & \text{for }\mathbf{x}\in T_{\srcidx}^- \\[1em]
\mathbf{0}, & \text{otherwise}
\end{matrix}
\right.,
\end{align*}
where \rereading{$\ell_{\srcidx}$ is the length of the edge shared by the triangle pair, and $A_{\srcidx}^+$ and $A_{\srcidx}^-$ are the areas of the triangles $T_{\srcidx}^+$ and $T_{\srcidx}^-$ associated with basis function $\srcidx$}.  $\boldsymbol{\rho}_{\srcidx}^+$ denotes the vector from the vertex of $T_{\srcidx}^+$ opposite the shared edge to $\mathbf{x}$, and $\boldsymbol{\rho}_{\srcidx}^-$ denotes the vector to the vertex of $T_{\srcidx}^-$ opposite the shared edge from $\mathbf{x}$.

These basis functions ensure that \reviewerOne{$\mathbf{J}_h$} is tangential to $S$ and has no component normal to the outer boundary of the triangle pair.  
Additionally, along \rereading{the shared edge of the triangle pair, the component of $\boldsymbol{\Lambda}_{\srcidx}(\mathbf{x})$ normal to that edge is unity.  Therefore, for a triangle edge shared by only two  triangles,} the component of \reviewerOne{$\mathbf{J}_h$} normal to that edge is $J_\srcidx$.  \reviewerTwo{Because the normal component is constant along the edge, the} solution is considered most accurate at the midpoint of the edge~\cite[pp.\ 155--156]{warnick_2008}.

Projecting~\eqref{eq:tan} onto $\boldsymbol{\Lambda}_{\testidx}(\mathbf{x})$ yields
\begin{align*}
\int_S \mathbf{E}^\mathcal{I}\cdot \boldsymbol{\Lambda}_{\testidx} dS =j\omega\int_S \mathbf{A}\cdot\boldsymbol{\Lambda}_{\testidx} dS + \int_S \nabla\Phi\cdot\boldsymbol{\Lambda}_{\testidx} dS, 
\end{align*}
which can be integrated by parts to obtain
\begin{align}
\int_S \mathbf{E}^\mathcal{I}\cdot \boldsymbol{\Lambda}_{\testidx} dS =j\omega\int_S \mathbf{A}\cdot\boldsymbol{\Lambda}_{\testidx} dS - \int_S \Phi \nabla\cdot\boldsymbol{\Lambda}_{\testidx} dS. 
\label{eq:proj}
\end{align}
Inserting~\eqref{eq:A} and~\eqref{eq:Phi} into \eqref{eq:proj} yields
\begin{align}
\int_S \mathbf{E}^\mathcal{I}\cdot \boldsymbol{\Lambda}_{\testidx} dS =&{}%
j\omega\mu \int_S \boldsymbol{\Lambda}_{\testidx}(\mathbf{x})\cdot\int_{S'} \mathbf{J}(\mathbf{x}')G(\mathbf{x},\mathbf{x}')dS'dS
{}-
\frac{j}{\epsilon\omega} \int_S \nabla\cdot\boldsymbol{\Lambda}_{\testidx}(\mathbf{x})\int_{S'} \nabla'\cdot\mathbf{J}(\mathbf{x}')G(\mathbf{x},\mathbf{x}')dS' dS.
\label{eq:proj_exp}
\end{align}
Substituting~\eqref{eq:J_approx} into~\eqref{eq:proj_exp}, we obtain \rereading{the discretized equation}
\begin{align}
\int_S \mathbf{E}^\mathcal{I}\cdot \boldsymbol{\Lambda}_{\testidx} dS ={}&%
j\omega\mu \sum_{\srcidx=1}^{\nbasis}  J_{\srcidx}\int_S \boldsymbol{\Lambda}_{\testidx}(\mathbf{x})\cdot\int_{S'} \boldsymbol{\Lambda}_{\srcidx}(\mathbf{x}')G(\mathbf{x},\mathbf{x}')dS'dS
\nonumber \\
&{}-%
\frac{j}{\epsilon\omega}\sum_{\srcidx=1}^{\nbasis} J_{\srcidx}\int_S \nabla\cdot\boldsymbol{\Lambda}_{\testidx}(\mathbf{x})\int_{S'} \nabla'\cdot\boldsymbol{\Lambda}_{\srcidx}(\mathbf{x}')G(\mathbf{x},\mathbf{x}')dS' dS.
\label{eq:proj_disc}
\end{align}
\reviewerOne{Letting $\mathbf{J}^h$ denote the vector of coefficients used to construct $\mathbf{J}_h$~\eqref{eq:J_approx}, \eqref{eq:proj_disc} can be written in matrix form as
\begin{align}
\mathbf{Z}\mathbf{J}^h = \mathbf{V},
\label{eq:zjv}
\end{align}}%
where
\begin{align*}
Z_{\testidx,\srcidx} &{}= j\omega\mu \int_S \boldsymbol{\Lambda}_{\testidx}(\mathbf{x})\cdot\int_{S'} \boldsymbol{\Lambda}_{\srcidx}(\mathbf{x}')G(\mathbf{x},\mathbf{x}')dS'dS - \frac{j}{\epsilon\omega} \int_S \nabla\cdot\boldsymbol{\Lambda}_{\testidx}(\mathbf{x})\int_{S'} \nabla'\cdot\boldsymbol{\Lambda}_{\srcidx}(\mathbf{x}')G(\mathbf{x},\mathbf{x}')dS' dS,
\\
\reviewerOne{J_{\srcidx}^h} &{}= J_{\srcidx},
\\
V_{\testidx} &{}=\int_S \mathbf{E}^\mathcal{I}\cdot \boldsymbol{\Lambda}_{\testidx} dS.
\end{align*}

\section{Manufactured Solutions}
\label{sec:mms}

Equation~\eqref{eq:proj_exp} can be written in terms of its residual \reviewerOne{as
%
$\mathbf{r}(\mathbf{J}) = \mathbf{0}$,
%
where
\begin{align*}
r_{\testidx}(\mathbf{J}) =&{}%
j\omega\mu \int_S \boldsymbol{\Lambda}_{\testidx}(\mathbf{x})\cdot\int_{S'} \mathbf{J}(\mathbf{x}')G(\mathbf{x},\mathbf{x}')dS'dS
{}-
\frac{j}{\epsilon\omega} \int_S \nabla\cdot\boldsymbol{\Lambda}_{\testidx}(\mathbf{x})\int_{S'} \nabla'\cdot\mathbf{J}(\mathbf{x}')G(\mathbf{x},\mathbf{x}')dS' dS -
\int_S \mathbf{E}^\mathcal{I}\cdot \boldsymbol{\Lambda}_{\testidx} dS.
\end{align*}
Similarly, its discretization~\eqref{eq:zjv} can be written as
\begin{align}
\mathbf{r}_h(\mathbf{J}_h) = \mathbf{0},
\label{eq:res_disc}
\end{align}
where $\mathbf{r}_h(\mathbf{J}_h) = \mathbf{Z}\mathbf{J}^h - \mathbf{V}$.}

The method of manufactured solutions modifies~\eqref{eq:res_disc} to be
\begin{align}
\mathbf{r}_h(\mathbf{J}_h) = \mathbf{r}(\mathbf{J}_\text{MS}),
\label{eq:mms}
\end{align}
where $\mathbf{J}_\text{MS}$ is the manufactured solution, and the generally nonzero $\mathbf{r}(\mathbf{J}_\text{MS})$ is computed analytically.  \reviewerOne{In~\eqref{eq:mms}, $\int_S \mathbf{E}^\mathcal{I}\cdot \boldsymbol{\Lambda}_{\testidx} dS$ appears on both sides, resulting in its cancellation.  However, instead of solving~\eqref{eq:mms}, we could equivalently solve~\eqref{eq:zjv} by inserting $\mathbf{J}_\text{MS}$ into~\eqref{eq:tan} to obtain a new incident electric field, which is projected onto $\boldsymbol{\Lambda}_{\testidx}(\mathbf{x})$ to obtain $\mathbf{V}$.

However, as described in the introduction, integrals containing the Green's function~\eqref{eq:G}, such as those appearing in this hypothetical proposal for $\mathbf{V}$ are not only unable to be computed analytically, but the singularity when $R\to 0$ complicates their accurate approximation, potentially contaminating convergence studies.

}%



\begin{figure}[!t]
\centering
\includegraphics[scale=.64,clip=true,trim=2.3in 0in 2.8in 0in]{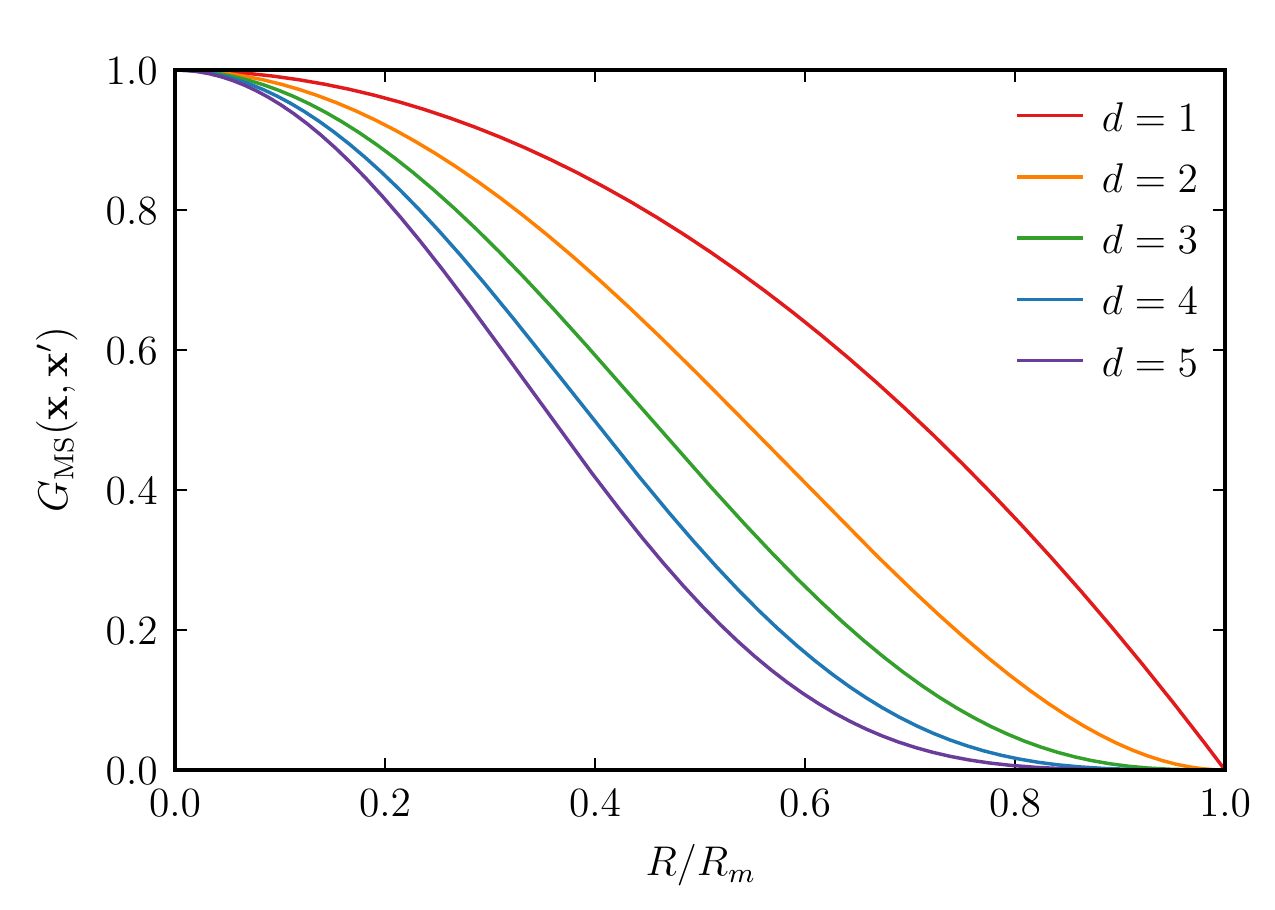}
\caption{$G_\text{MS}$~\eqref{eq:G_mms} for different values of $d$.}
\vskip-\dp\strutbox
\label{fig:G_MS}
\end{figure}

Nonetheless, we can still verify the order of accuracy of the solution-discretization error, Error Source~\eqref{err:sol_disc}, by also manufacturing the Green's function, using the form
\begin{align}
G_\text{MS}(\mathbf{x},\mathbf{x}') = \left(1 - \frac{R^2}{R_m^2}\right)^d,
\label{eq:G_mms}
\end{align}
where $R_m=\max_{\mathbf{x},\mathbf{x}'\in S} R$ is the maximum possible distance between two points on $S$, and $d\in\mathbb{N}$.  A plot of~\eqref{eq:G_mms} is shown in Figure~\ref{fig:G_MS} for multiple values of $d$.  The form of~\eqref{eq:G_mms} is chosen for two reasons: 1) the even powers of $R$ permit the integrals in~\eqref{eq:proj_disc} and
\reviewerOne{$\mathbf{V}$
to be computed analytically for many choices of $\mathbf{J}_\text{MS}$, avoiding contamination from additional error}, and 2) $G_\text{MS}$ increases when $R$ decreases, as with the actual Green's function~\eqref{eq:G}.
Introducing the terms $\mathbf{Z}_\text{MS}$ and $\mathbf{V}_\text{MS}$ to account for $G_\text{MS}$, we write the resulting system of equations as
\begin{align}
\mathbf{r}_\text{MS} = \mathbf{Z}_\text{MS}\mathbf{J}^h - \mathbf{V}_\text{MS} = \mathbf{0},
\label{eq:zjv_mms}
\end{align}
where
\reviewerOne{
\begin{align}
Z_{\text{MS}_{\testidx,\srcidx}} &{}= j\omega\mu \int_S \boldsymbol{\Lambda}_{\testidx}(\mathbf{x})\cdot\int_{S'} \boldsymbol{\Lambda}_{\srcidx}(\mathbf{x}')G_\text{MS}(\mathbf{x},\mathbf{x}')dS'dS - \frac{j}{\epsilon\omega} \int_S \nabla\cdot\boldsymbol{\Lambda}_{\testidx}(\mathbf{x})\int_{S'} \nabla'\cdot\boldsymbol{\Lambda}_{\srcidx}(\mathbf{x}')G_\text{MS}(\mathbf{x},\mathbf{x}')dS' dS, \nonumber \\
V_{\text{MS}_{\testidx}} &{}=\int_S \mathbf{E}_\text{MS}^\mathcal{I}\cdot \boldsymbol{\Lambda}_{\testidx} dS, \label{eq:V_mms}
\end{align}}
and
\begin{align}
\mathbf{E}_\text{MS}^\mathcal{I}(\mathbf{x}) &{}= \frac{j}{\omega\epsilon} \int_{S'}\left( k^2\mathbf{J}_\text{MS} (\mathbf{x}')G_\text{MS}(\mathbf{x},\mathbf{x}') +\nabla'\cdot\mathbf{J}_\text{MS}(\mathbf{x}')\nabla G_\text{MS}(\mathbf{x},\mathbf{x}')\right)dS'. 
\label{eq:Ei_mms}
\end{align}

Solving~\eqref{eq:zjv_mms} for $\mathbf{J}^h$ enables us to compute the discretization error 
\reviewerTwo{%
\begin{align}
\mathbf{e}_n = \mathbf{J}^h - \mathbf{J}_n,
\label{eq:error}
\end{align}%
where $J_{n_\srcidx}$ denotes the component of $\mathbf{J}_\text{MS}$ flowing from $T_\srcidx^+$ to $T_\srcidx^-$.  The norm of~\eqref{eq:error} has the property $\|\mathbf{e}_n\|\le C h^p$, where
$C$ is a function of the solution derivatives}, $h$ is representative of the mesh size, and $p$ is the order of accuracy.  By performing a mesh-convergence study of \reviewerTwo{the norm of} the discretization error, we can ensure the proper order of accuracy is obtained.  For the RWG basis functions, the expectation is second-order accuracy $(p=2)$.

Despite the benefit of yielding analytically integrable integrals in~\eqref{eq:zjv_mms}, $G_\text{MS}$ makes $\mathbf{Z}_\text{MS}$ practically singular, admitting multiple solutions $\mathbf{J}^h$ in~\eqref{eq:zjv_mms}.  Therefore, we choose the $\mathbf{J}^h$ that \reviewerOne{both yields $\mathbf{J}_h$ closest} to $\mathbf{J}_\text{MS}$ and satisfies~\eqref{eq:zjv_mms}.

To accomplish this, we perform a pivoted QR factorization of $\mathbf{Z}_\text{MS}^H$, such that
\begin{align}
\mathbf{Z}_\text{MS}^H\mathbf{P} = \left[\mathbf{Q}_1,\,\mathbf{Q}_2\right]\left[\begin{matrix} \mathbf{R}_1 \\ \mathbf{0} \end{matrix}\right] =  \mathbf{Q}_1 \mathbf{R}_1,
\label{eq:qr}
\end{align}
where $\mathbf{Z}_\text{MS}\in\mathbb{C}^{\nbasis\times\nbasis}$, $\mathbf{Q}_1\in\mathbb{C}^{\nbasis\times\rank}$, $\mathbf{Q}_2\in\mathbb{C}^{\nbasis\times(\nbasis-\rank)}$, and $\mathbf{R}_1\in\mathbb{C}^{\rank\times\nbasis}$.  $\mathbf{P}$ is a permutation matrix, and \rereading{the superscript $H$ denotes conjugated transposition}.
Numerically, the pivoting facilitates the determination of the rank $\rank\le\nbasis$ of $\mathbf{Z}_\text{MS}$.  We can express $\mathbf{J}^h$ in terms of the basis $\mathbf{Q}$:
\begin{align}
\mathbf{J}^h = \mathbf{Q}_1 \mathbf{u} + \mathbf{Q}_2 \mathbf{v},
\label{eq:J_qbasis}
\end{align}
where $\mathbf{u}\in\mathbb{C}^{\rank}$ is the vector of coefficients that satisfies~\eqref{eq:zjv_mms} and $\mathbf{v}\in\mathbb{C}^{\nbasis-\rank}$ is the vector of coefficients that brings $\mathbf{J}_h$ closest to $\mathbf{J}_\text{MS}$, given $\mathbf{u}$.
Substituting~\eqref{eq:qr} and~\eqref{eq:J_qbasis} into~\eqref{eq:zjv_mms} and noting that $\mathbf{Q}$ is unitary, $\mathbf{u}$ can be computed from 
\begin{align}
\mathbf{R}_1^H\mathbf{u} = \mathbf{P}^T\mathbf{V}_\text{MS}.
\label{eq:u}
\end{align}
From $\mathbf{u}$ satisfying~\eqref{eq:u}, $\mathbf{J}^h$ satisfies~\eqref{eq:zjv_mms}.  With $\mathbf{u}$ known, $\mathbf{v}$ remains to be determined.  Let
\begin{align}
\bar{\mathbf{J}}^h = \mathbf{Q}_1 \mathbf{u}.
\label{eq:Jbar}
\end{align}
We solve for $\mathbf{v}$ by minimizing 
\begin{align}
\rereading{\|\mathbf{e}_n\|_2^2} = \left(\mathbf{J}^h-\mathbf{J}_n\right)^H\left(\mathbf{J}^h-\mathbf{J}_n\right).
\label{eq:error_L2}
\end{align}
Substituting~\eqref{eq:J_qbasis} into~\eqref{eq:error_L2} and using~\eqref{eq:Jbar}, \eqref{eq:error_L2} can be written as
\begin{align}
\rereading{\|\mathbf{e}_n\|_2^2 = \left[\mathbf{Q}_2\mathbf{v} - \left(\mathbf{J}_n-\bar{\mathbf{J}}^h\right)\right]^H\left[\mathbf{Q}_2\mathbf{v} - \left(\mathbf{J}_n-\bar{\mathbf{J}}^h\right)\right]}.
%
\label{eq:quadratic}
\end{align}
Equation~\eqref{eq:quadratic} is quadratic, and is minimized when
\begin{align*}
\mathbf{v} = \mathbf{Q}_2^H\left(\mathbf{J}_n-\bar{\mathbf{J}}^h\right).
\end{align*}
Therefore, the expression for $\mathbf{J}^h$ is
\begin{align}
\mathbf{J}^h = \bar{\mathbf{J}}^h + \mathbf{Q}_2\mathbf{Q}_2^H\left(\mathbf{J}_n-\bar{\mathbf{J}}^h\right).
\label{eq:Jh_sol}
\end{align}
Equation~\eqref{eq:Jh_sol} provides us with the solution to~\eqref{eq:zjv_mms} that is closest to $\mathbf{J}_\text{MS}$.

\section{Numerical Examples} 
\label{sec:results}

In this section, we demonstrate the effectiveness of the approach described in Section~\ref{sec:mms}.
As stated in the introduction, Error Source~\ref{err:dom_disc} is outside of the scope of this work.  Therefore, we restrict our domain to planar regions.  Specifically, we consider two unit-square flat plates, as shown in Figure~\ref{fig:two_squares}, with one rotated out of the plane of the other by angle $\theta$.

\begin{figure}[!t]
\centering
\begin{subfigure}[b]{\textwidth}
\centering
\includegraphics[scale=.28,clip=true,trim=0in 0in 0in 0in]{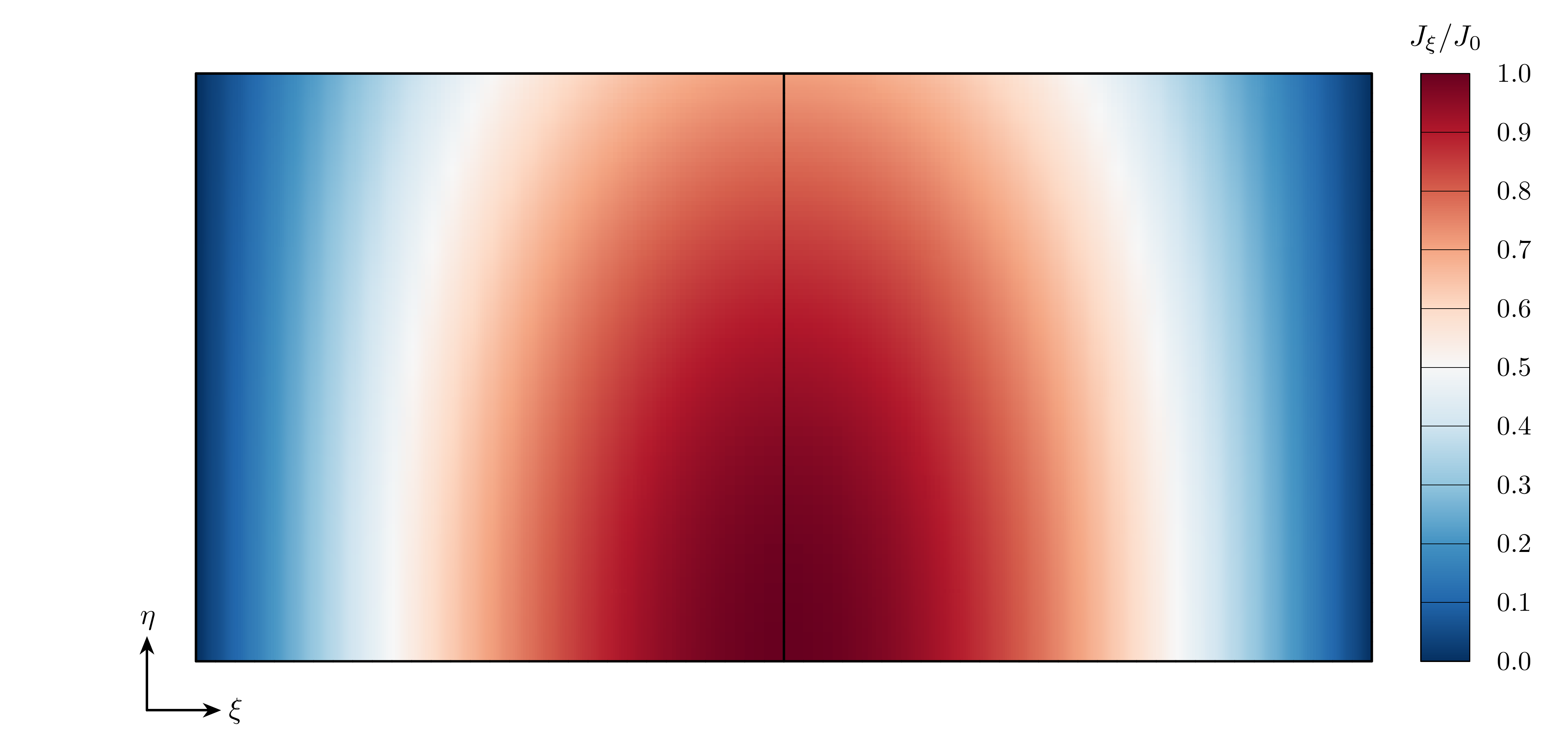}
\caption{\strut$J_\xi$}
\label{fig:Jxi}
\end{subfigure}

\begin{subfigure}[b]{\textwidth}
\centering
\includegraphics[scale=.28,clip=true,trim=0in 0in 0in 0in]{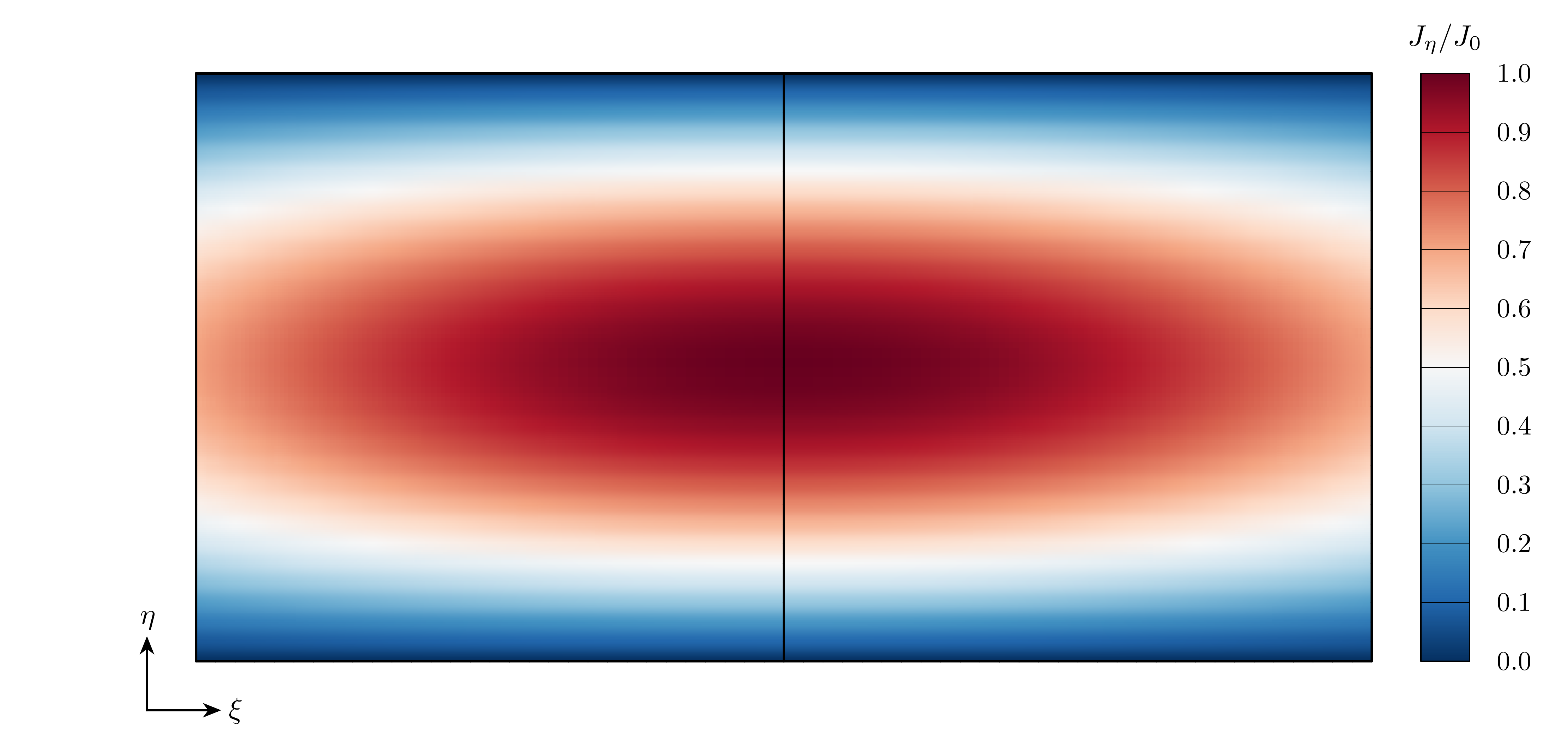}
\caption{\strut$J_\eta$}
\label{fig:Jeta}
\end{subfigure}
\caption{\strut Manufactured surface current $\mathbf{J}_\text{MS}$.}
\vskip-\dp\strutbox
\label{fig:J_MS}
\end{figure}

We manufacture the surface current $\mathbf{J}_\text{MS}(\mathbf{x}) = \{J_\xi(\boldsymbol{\xi}),\,J_\eta(\boldsymbol{\xi})\}$ using sinusoidal functions:
\begin{align}
J_\xi (\boldsymbol{\xi}) &{}= J_0 \cos(\pi \xi/2)\cos(\pi \eta/4), \label{eq:Jxi}\\
J_\eta(\boldsymbol{\xi}) &{}= J_0 \cos(\pi \xi/4)\sin(\pi \eta), \label{eq:Jeta}
\end{align}
where $J_0=1$ A/m and the plate-fixed coordinate system $\boldsymbol{\xi}(\mathbf{x};\theta)$ is given by
\reviewerOne{%
\begin{align*}
\xi(\mathbf{x};\theta) &{}= \frac{1}{L_0}\left\{\begin{matrix}
x, & \text{for } x\le 0 \text{ m} \\
x\cos\theta + z\sin\theta, & \text{for } x>0 \text{ m}
\end{matrix}\right.,
\\
\eta(\mathbf{x}) &{}= y/L_0,
\end{align*}
where $L_0=1$ m.}
At the edges of the domain, the normal component of $\mathbf{J}_\text{MS}(\mathbf{x})$ is zero, satisfying the boundary conditions.  Figures~\ref{fig:Jxi} and~\ref{fig:Jeta} provide plots of~\eqref{eq:Jxi} and~\eqref{eq:Jeta}.

\newsavebox{\twosquares}%
\sbox{\twosquares}{%
\newcommand*\projA{-20}
\newcommand*\projB{ 35}
%
\definecolor{darkred} {RGB}{227,26,28}%
\definecolor{darkorange} {RGB}{255,127,0}%
\definecolor{lightred} {RGB}{243.8,163.4,164.2}%
\definecolor{darkblue}  {RGB}{ 31,120,180}%
\definecolor{lightblue}  {RGB}{165.4,201,225}
\definecolor{darkgreen} {RGB}{ 51,160, 44}%
\definecolor{lightgreen}{RGB}{173.4,217,170.6}
\def\xangle{100}
\def\yangle{60}
\begin{tikzpicture}[scale=0.6,
x={({ sin(\xangle)*1cm},{ cos(\xangle)*1cm})}, 
y={({.9*sin(\yangle)*1cm},{.9*cos(\yangle)*1cm})}, 
z={(0cm,1cm)}
    ]   

   \def\Deltaz{0};    
   \def\Deltay{0};    
   \def\h{6};         
   \def\xl{-6};       
   \def\xr{0};        
   \def\meas{1};      
   \def\mw{.25};      
   \def\thetaval{45}; 
   \def\arcrad{1};    
   \def\ep{.1}        
   \def\ymax{7}      
   
   \useasboundingbox (-11cm,-3cm) rectangle (7cm,9cm);
   
   \begin{scope}[canvas is xy plane at z=0,transform shape]
   \draw[line cap=round] (0,0) -- (-7+\ep, 0);
   \draw[line cap=round] (0,0) -- (0,\ymax-\ep);
   \pgflowlevelsynccm
   \draw[-{Stealth[scale=2.0]},line cap=round] (0,0) to (-7, 0) node[left,scale=2,dash pattern=on 0cm off 100cm] {$y=\eta$};
   \draw[-{Stealth[scale=2.0]},line cap=round] (0,0) to (0,\ymax) node[above,scale=2,dash pattern=on 0cm off 100cm] {$x$};
   \end{scope}

   \begin{scope}[canvas is yz plane at x=0,transform shape]
      \draw[line cap=round,gray] (\arcrad,0) arc (0:\thetaval:\arcrad);
      \node[scale=2] at ({1.5*\arcrad*cos(\thetaval/2)},{1.5*\arcrad*sin(\thetaval/2)}) {$\theta$};
      \pgflowlevelsynccm
      
   \end{scope}

   \begin{scope}[canvas is plane={O(0,0,0)x(1,0,0)y(0,{cos(\thetaval)},{sin(\thetaval)})}]
      \draw[line cap=round,dashed,gray] (0,0) -- (0,7-\ep);
      
      \pgflowlevelsynccm
      \draw[-{Stealth[scale=2.0,fill=gray,gray]},line cap=round,dash pattern=on 0cm off 100cm,text=gray] (0,0) to (0,7) node[above,scale=2] {$\xi$};
      
      \draw[thick,draw=darkgreen,fill=lightgreen,fill opacity=.5,draw opacity=1] (\xr,0) -- (\xr,\h) -- (\xl,\h) -- (\xl,0) -- cycle;

      \pgflowlevelsynccm
      \draw[-{Stealth[scale=2.0,fill=gray,gray]},line cap=round,dash pattern=on 0cm off 100cm,darkgreen] (0,0) to (0,7); 
      
      \node[transform shape,scale=2,text=darkgreen,anchor=west] at (\xr,\h) {$(1,0)$};
      \node[transform shape,scale=2,text=darkgreen,anchor=east] at (\xl,\h) {$(1,1)$};
      \draw[draw=darkgreen,fill=darkgreen] (\xr,\h) circle (.1);
      \draw[draw=darkgreen,fill=darkgreen] (\xl,\h) circle (.1);
   \end{scope}

   \begin{scope}[canvas is xz plane at y=0]
   \draw[line cap=round] (0,0) -- (0,8-\ep);
   \pgflowlevelsynccm
   \draw[-{Stealth[scale=2.0]},line cap=round,dash pattern=on 0cm off 100cm] (0,0) to (0,8) node[above,scale=2] {$z$};
   \end{scope}

   \begin{scope}[canvas is xy plane at z=0]
      \draw[thick,darkblue,fill=lightblue,fill opacity=.5,draw opacity=1] (\xr,-\h) -- (\xr,0) -- (\xl,0) -- (\xl,-\h) -- cycle;

      \node[transform shape,scale=2,text=darkblue,anchor=west] at (\xr,-\h) {$(-1,0)$};
      \node[transform shape,scale=2,text=darkblue,anchor=east] at (\xl,-\h) {$(-1,1)$};
      %
      \draw[draw=darkblue,fill=darkblue] (\xr,-\h) circle (.1);
      \draw[draw=darkblue,fill=darkblue] (\xl,-\h) circle (.1);
   \end{scope}

\end{tikzpicture}%
}%
\begin{figure}[!b]
\centering
\usebox{\twosquares}%
\caption{Computational domain consisting of two unit-square plates.  \reviewerOne{Coordinates are expressed in the plate-fixed coordinate system $\boldsymbol{\xi}(\mathbf{x};\theta)$.}}
\label{fig:two_squares}
\end{figure}

\begin{figure}[!t]
\centering
\begin{subfigure}[b]{\textwidth}
\centering
\includegraphics[scale=.28,clip=true,trim=0in 0in 0in 0in]{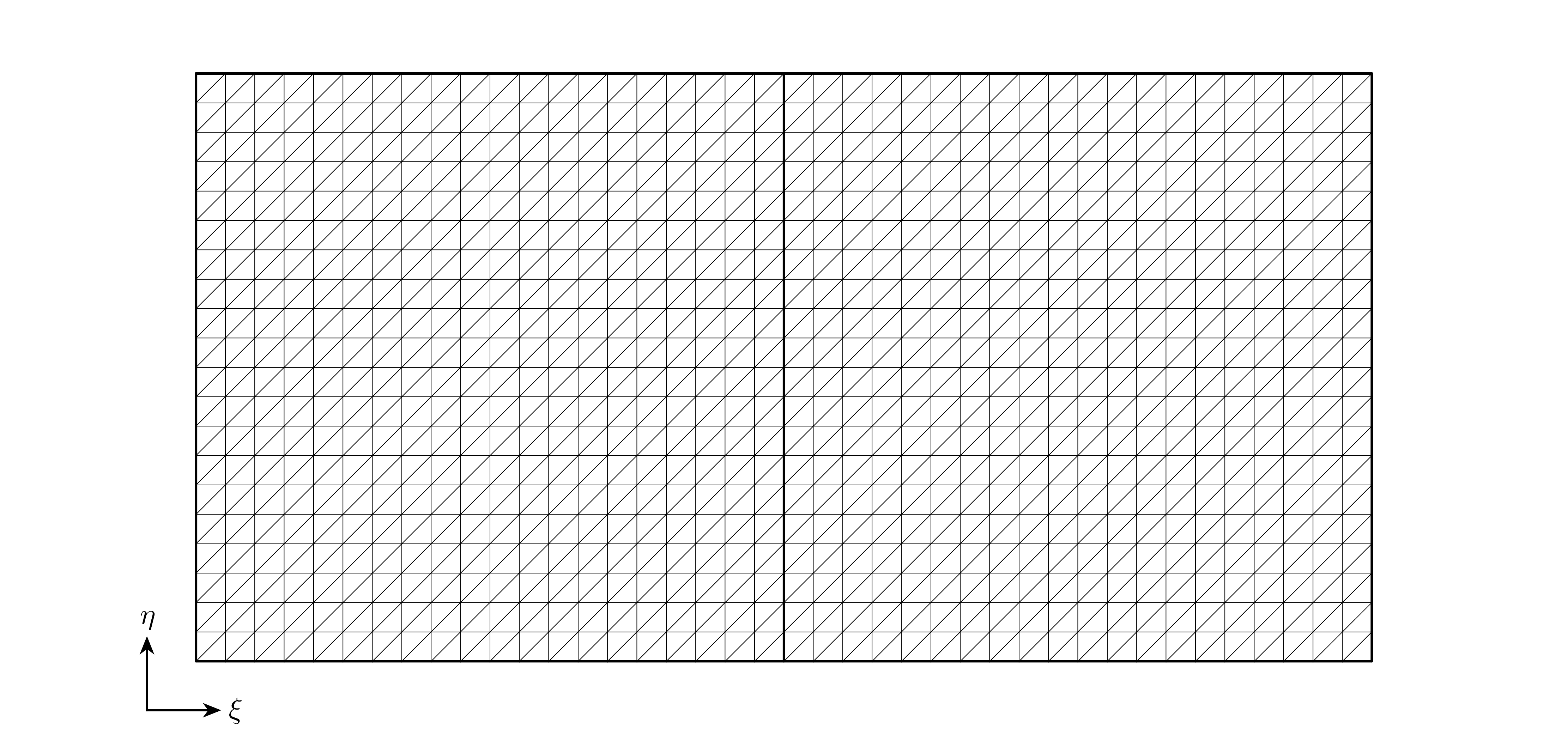}
\caption{\strut Uniform}
\label{fig:uniform}
\end{subfigure}

\begin{subfigure}[b]{\textwidth}
\centering
\includegraphics[scale=.28,clip=true,trim=0in 0in 0in 0in]{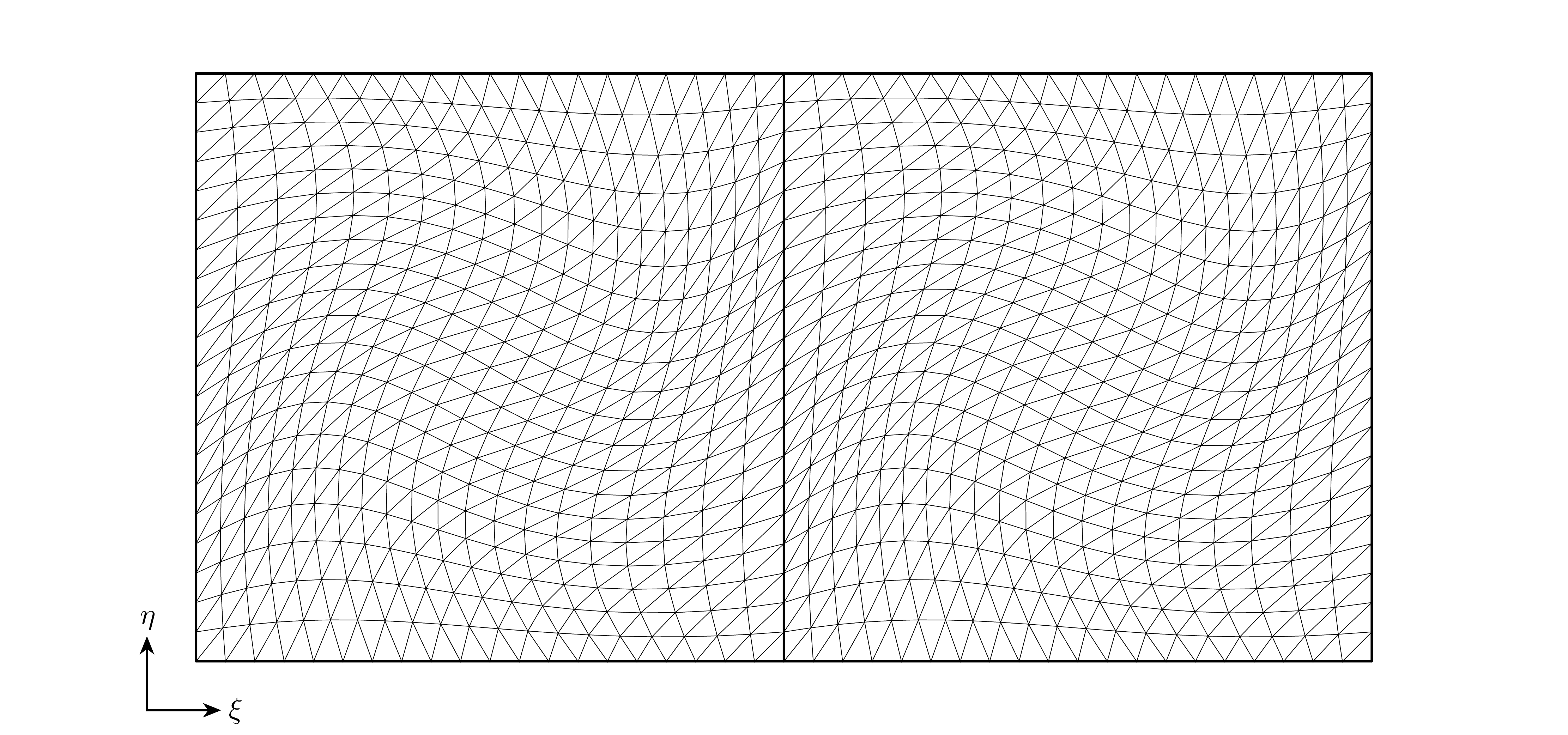}
\caption{\strut Twisted}
\label{fig:twisted}
\end{subfigure}
\caption{\strut Two different types of meshes, shown with $\ntriangles=1600$.}
\vskip-\dp\strutbox
\label{fig:mesh}
\end{figure}

We consider two types of meshes: a uniform mesh and a twisted mesh, examples of which are shown in Figures~\ref{fig:uniform} and~\ref{fig:twisted} with $\ntriangles=1600$ triangles, as well as $d=1$ and $d=2$ in the manufactured Green's function $G_\text{MS}$~\eqref{eq:G_mms}.  The twisted mesh is obtained by transforming the uniform mesh, using the transformation provided in Reference~\cite{freno_2021}.

We measure the $L^\infty$-norm of the error $\mathbf{e}_n$~\eqref{eq:error}:
\begin{align}
\varepsilon = \left\|\mathbf{e}_n\right\|_\infty = \max|\mathbf{J}^h-\mathbf{J}_n|
\label{eq:error_norm}
\end{align}
to determine the maximum error in the surface current normal to the interior edges at the edge midpoints.

Finally, we account for potential disparities in the magnitudes of the contributions to $\mathbf{Z}_\text{MS}$ from $\mathbf{A}$ and $\Phi$:  
\begin{align*}
\mathbf{Z}_\text{MS} = \mathbf{Z}^\mathbf{A} + \mathbf{Z}^\Phi,
\end{align*}
where
\begin{align*}
Z_{\testidx,\srcidx}^\mathbf{A} &{}= \frac{j k^2}{\epsilon\omega} \int_S \boldsymbol{\Lambda}_{\testidx}(\mathbf{x})\cdot\int_{S'} \boldsymbol{\Lambda}_{\srcidx}(\mathbf{x}')G_\text{MS}(\mathbf{x},\mathbf{x}')dS'dS ,
\\
Z_{\testidx,\srcidx}^\Phi &{}=  - \frac{j}{\epsilon\omega} \int_S \nabla\cdot\boldsymbol{\Lambda}_{\testidx}(\mathbf{x})\int_{S'} \nabla'\cdot\boldsymbol{\Lambda}_{\srcidx}(\mathbf{x}')G_\text{MS}(\mathbf{x},\mathbf{x}')dS' dS.
\end{align*}
We consider the contributions $\mathbf{Z}^\mathbf{A}$ and $\mathbf{Z}^\Phi$ together and separately, with $\epsilon=1$ F/m and $\mu=1$ H/m.  When we consider the contributions together, we set $k=1$ m$^{-1}$ for $\mathbf{Z}_\text{MS}$.  When we consider them separately, we are effectively taking the limits as $k\to\infty$ for $\mathbf{Z}^\mathbf{A}$ and $k\to 0$ for $\mathbf{Z}^\Phi$.  We adjust $\mathbf{E}_\text{MS}^\mathcal{I}$~\eqref{eq:Ei_mms} accordingly.

\subsection{Correct Implementation} 

\begin{figure}[!t]
\centering
\begin{subfigure}[b]{.49\textwidth}
\includegraphics[scale=.64,clip=true,trim=2.3in 0in 2.8in 0in]{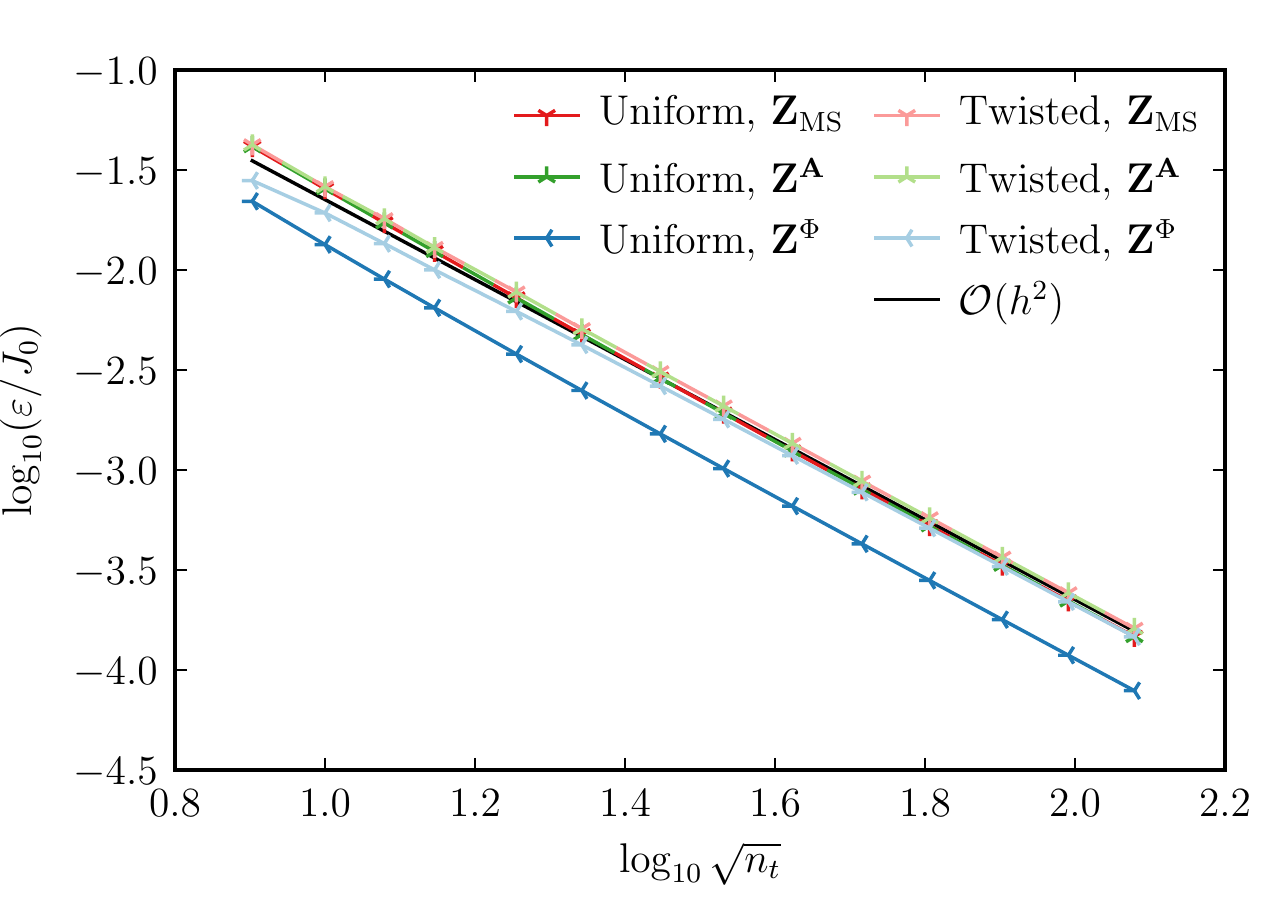}
\caption{$\theta=0^\circ$}
\end{subfigure}
\hspace{0.25em}
\begin{subfigure}[b]{.49\textwidth}
\includegraphics[scale=.64,clip=true,trim=2.3in 0in 2.8in 0in]{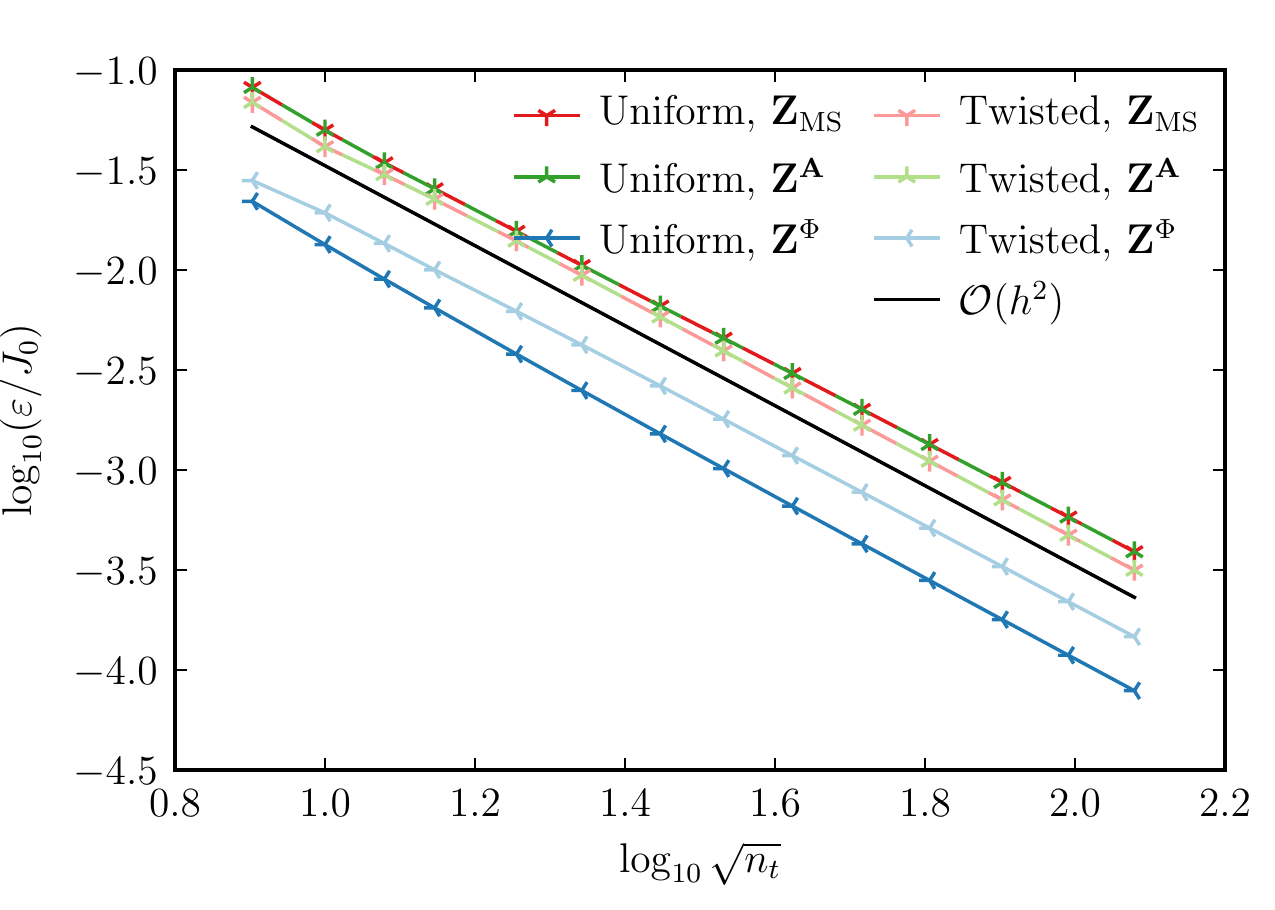}
\caption{$\theta=45^\circ$}
\label{fig:G1:45}
\end{subfigure}
\\
\begin{subfigure}[b]{.49\textwidth}
\includegraphics[scale=.64,clip=true,trim=2.3in 0in 2.8in 0in]{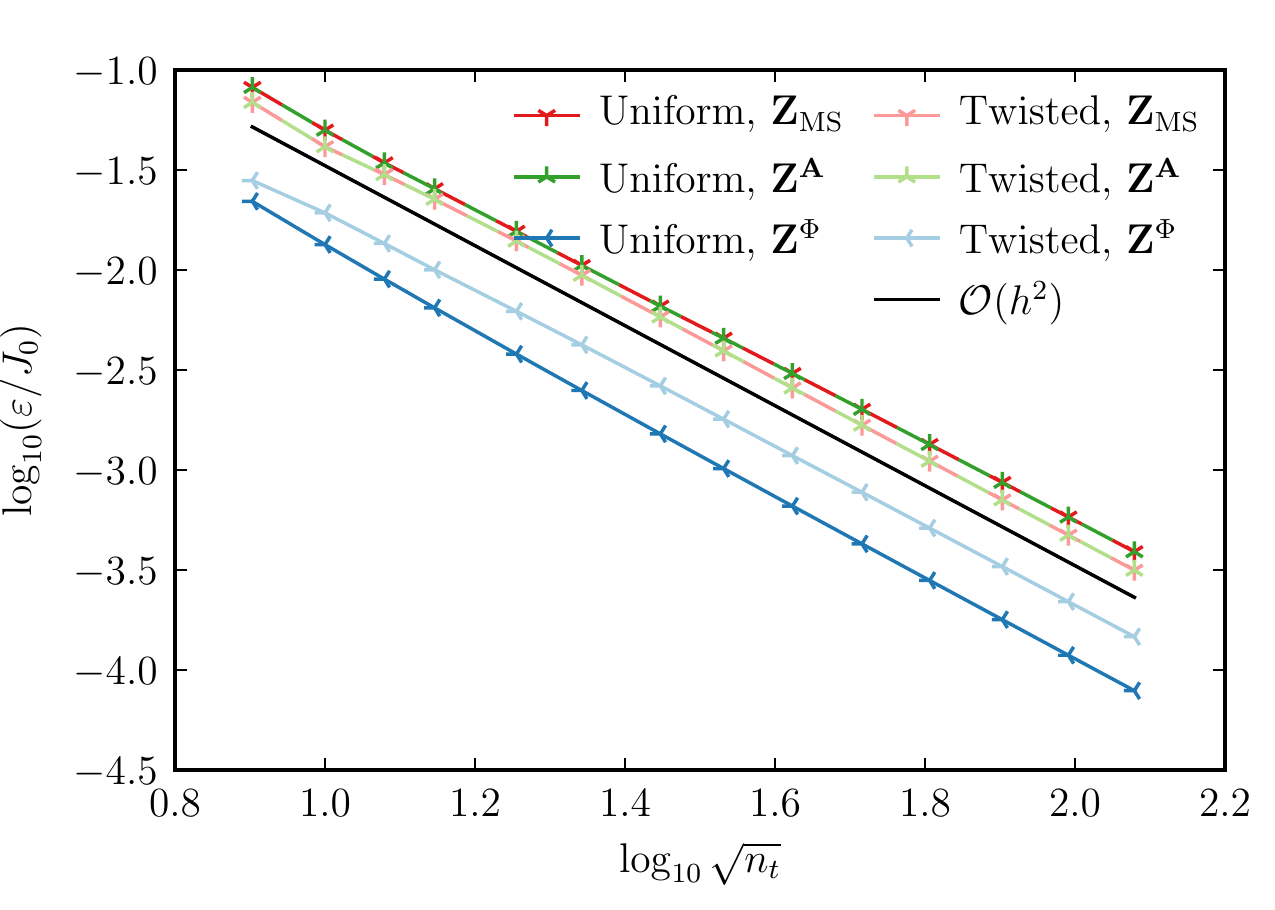}
\caption{$\theta=90^\circ$}
\end{subfigure}
\hspace{0.25em}
\begin{subfigure}[b]{.49\textwidth}
\includegraphics[scale=.64,clip=true,trim=2.3in 0in 2.8in 0in]{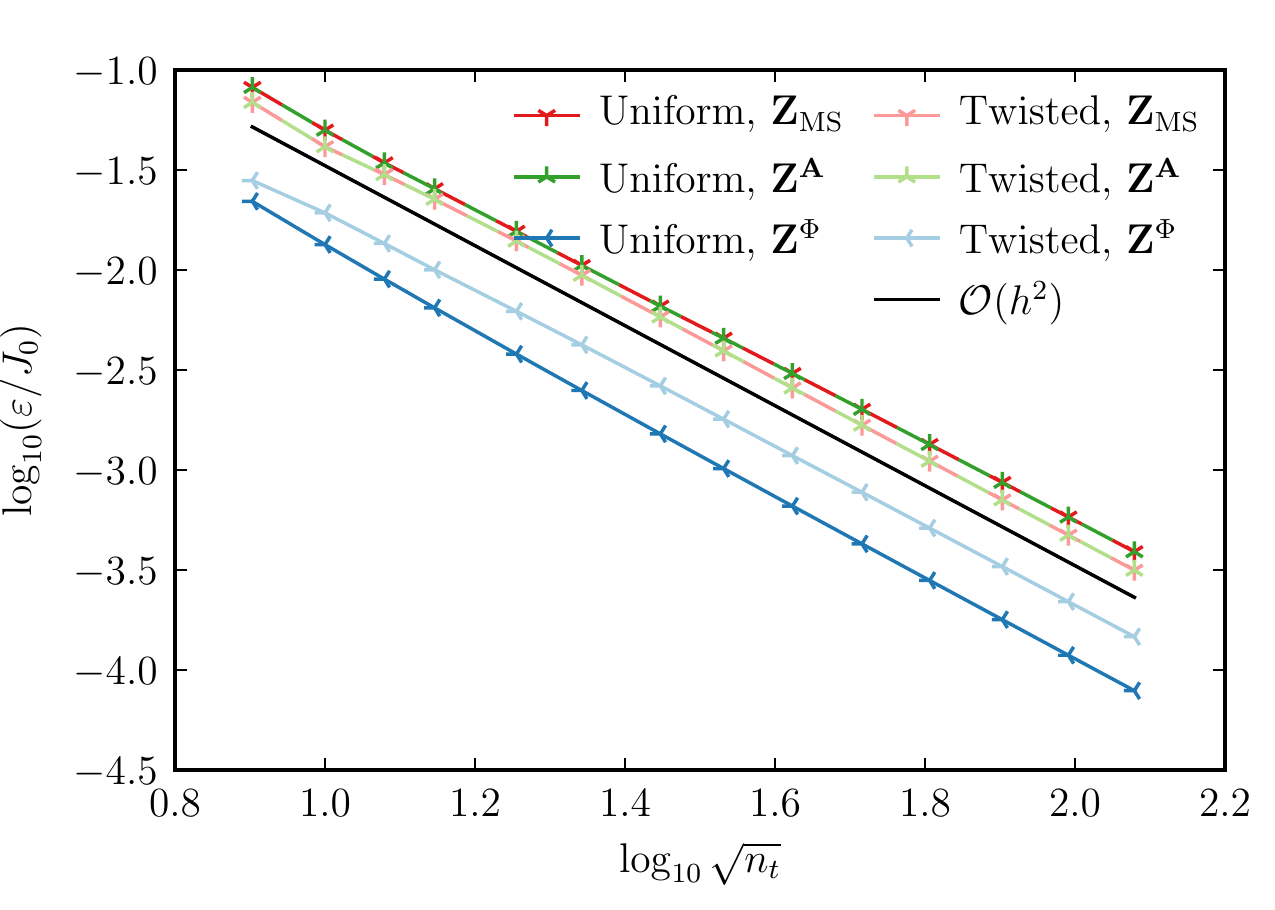}
\caption{$\theta=135^\circ$}
\end{subfigure}
\caption{Error $\varepsilon$~\eqref{eq:error_norm} for $d=1$ in $G_\text{MS}$~\eqref{eq:G_mms}.}
\vskip-\dp\strutbox
\vskip10em
\label{fig:G1}
\end{figure}

We first consider the case of $d=1$ in~\eqref{eq:G_mms}.  For this choice of $G_\text{MS}$, $\mathbf{E}_\text{MS}^\mathcal{I}$~\eqref{eq:Ei_mms} can be computed analytically to yield a polynomial expression of degree two, such that the integrand of~\eqref{eq:V_mms} is a polynomial of degree three, which can be integrated exactly using four triangle quadrature points for polynomials~\cite{lyness_1975}.
Figure~\ref{fig:G1} shows plots of $\varepsilon$~\eqref{eq:error_norm} with respect to $\ntriangles$ for four $\theta$ values and the two meshes.  The plots indicate that the solution is second-order accurate, as expected.  Table~\ref{tab:G1} shows the maximum rank, $\max\rank$, of~\eqref{eq:zjv_mms} for each of the simulations, across the different mesh sizes.  These low values emphasize how poorly conditioned $\mathbf{Z}_\text{MS}$ is.

\begin{figure}[!t]
\centering
\begin{subfigure}[b]{.49\textwidth}
\includegraphics[scale=.64,clip=true,trim=2.3in 0in 2.8in 0in]{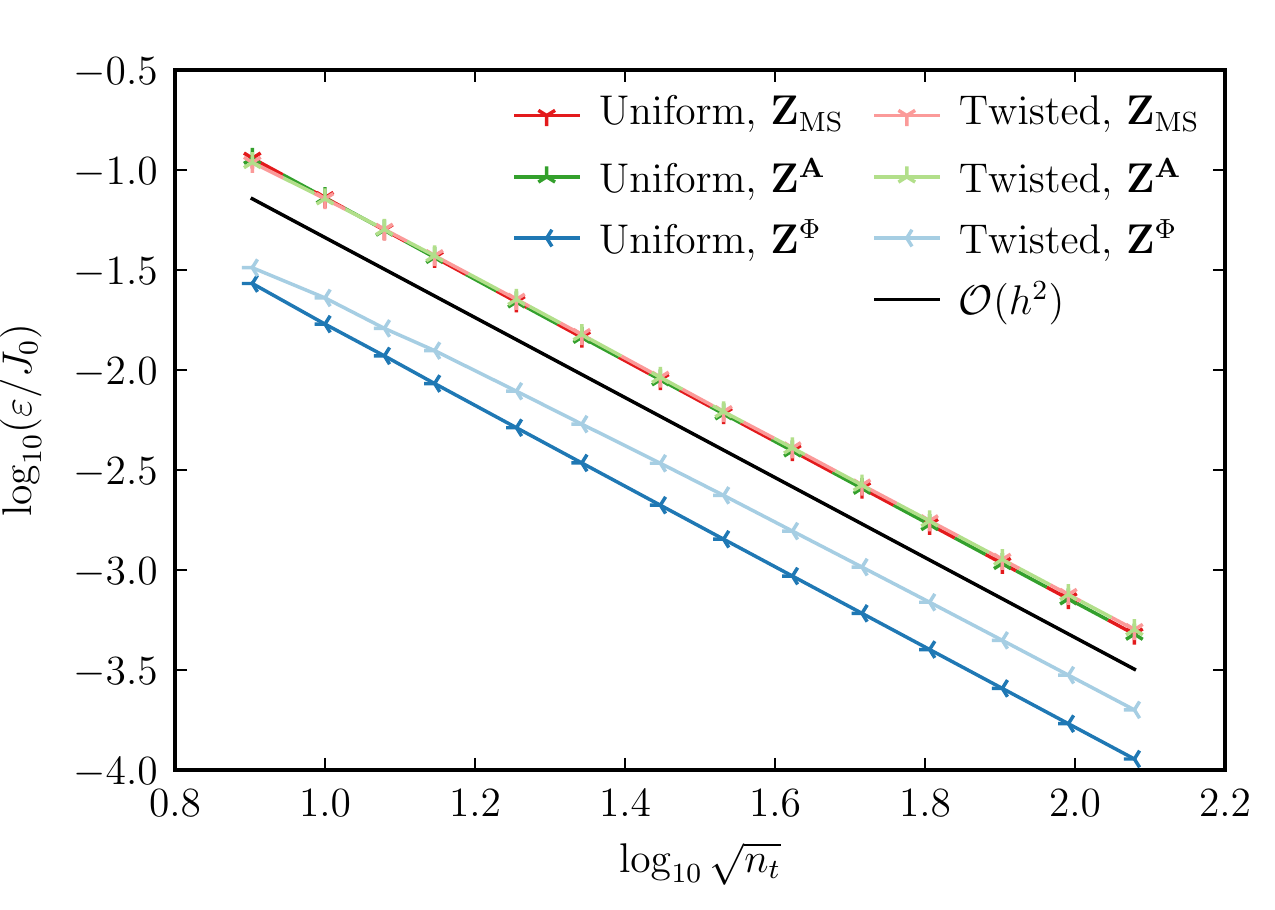}
\caption{$\theta=0^\circ$}
\end{subfigure}
\hspace{0.25em}
\begin{subfigure}[b]{.49\textwidth}
\includegraphics[scale=.64,clip=true,trim=2.3in 0in 2.8in 0in]{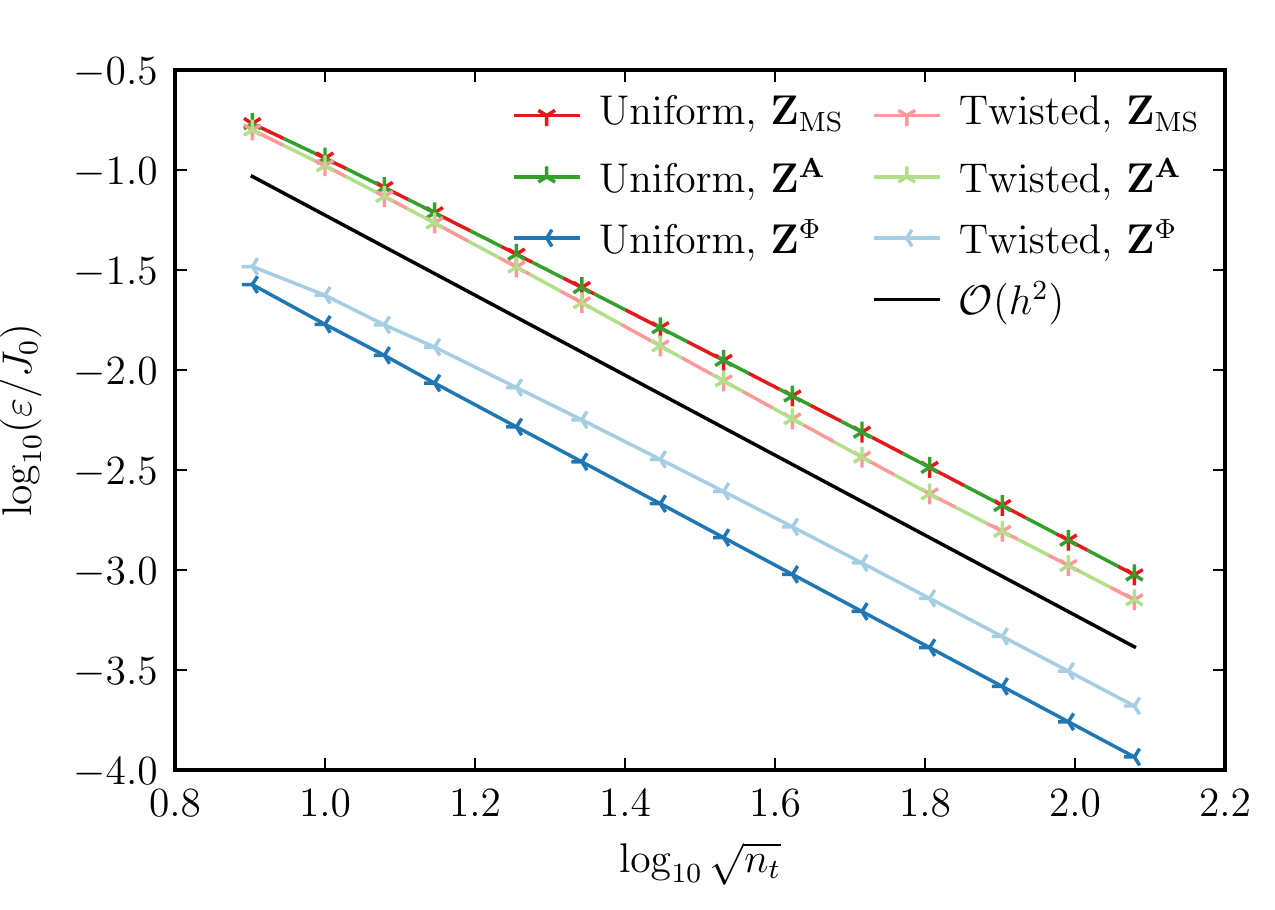}
\caption{$\theta=45^\circ$}
\end{subfigure}
\\
\begin{subfigure}[b]{.49\textwidth}
\includegraphics[scale=.64,clip=true,trim=2.3in 0in 2.8in 0in]{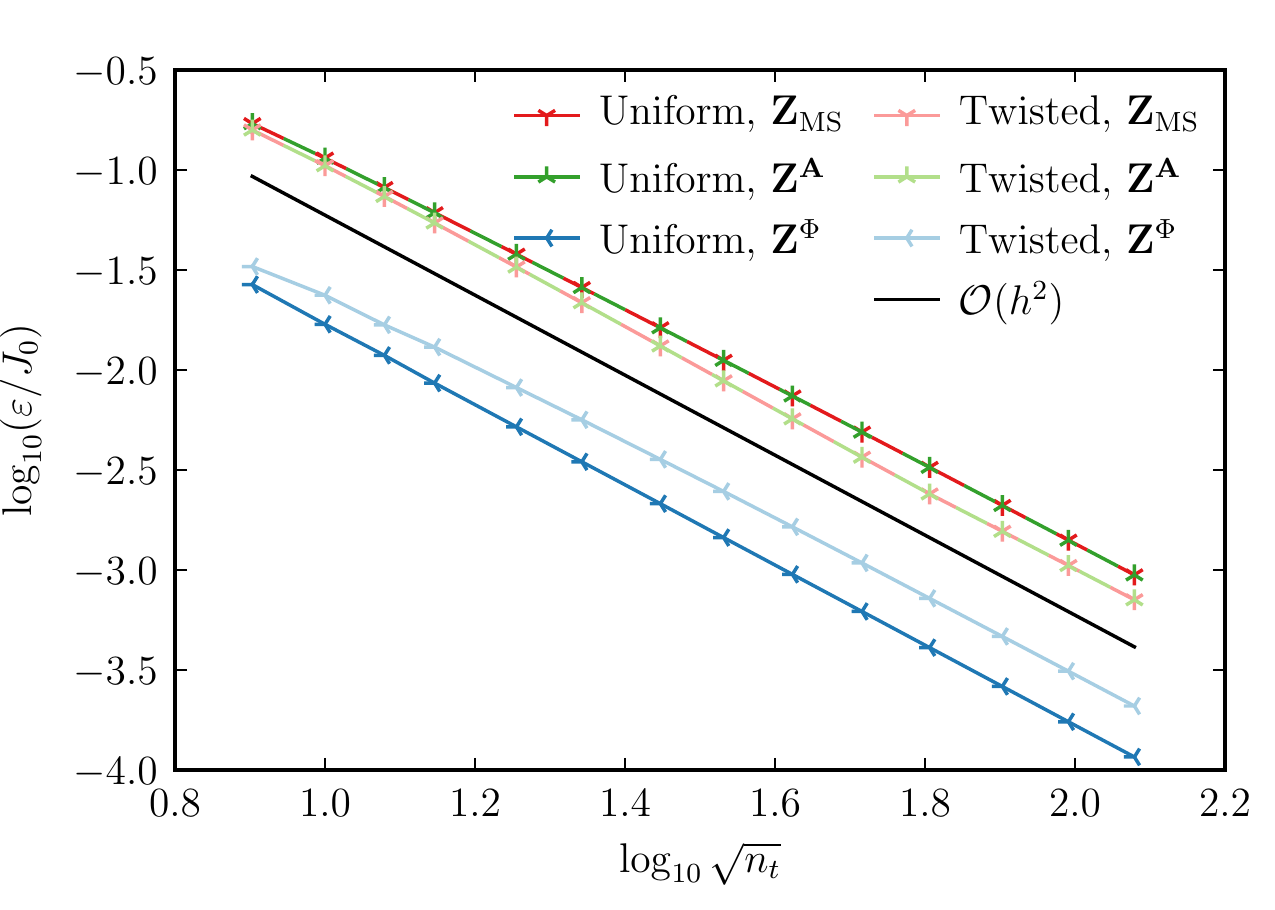}
\caption{$\theta=90^\circ$}
\end{subfigure}
\hspace{0.25em}
\begin{subfigure}[b]{.49\textwidth}
\includegraphics[scale=.64,clip=true,trim=2.3in 0in 2.8in 0in]{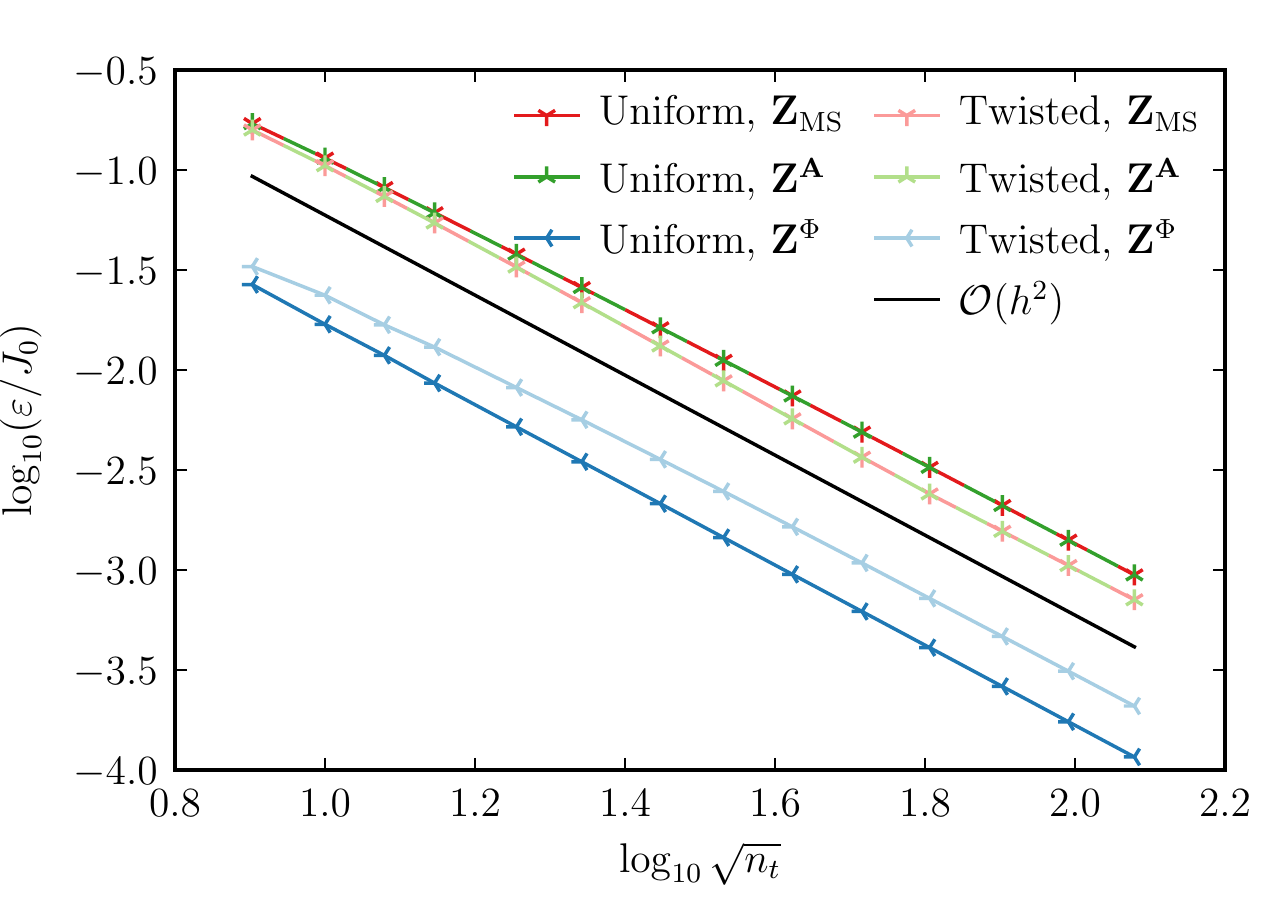}
\caption{$\theta=135^\circ$}
\end{subfigure}
\caption{Error $\varepsilon$~\eqref{eq:error_norm} for $d=2$ in $G_\text{MS}$~\eqref{eq:G_mms}.}
\vskip-\dp\strutbox
\vskip10em
\label{fig:G2}
\end{figure}

Next, we consider the case of $d=2$ in~\eqref{eq:G_mms}.  For this choice of $G_\text{MS}$, $\mathbf{E}_\text{MS}^\mathcal{I}$~\eqref{eq:Ei_mms} can be computed analytically to yield a polynomial expression of degree four, such that the integrand of~\eqref{eq:V_mms} is a polynomial of degree five, which can be integrated exactly using seven triangle quadrature points for polynomials~\cite{lyness_1975}.
Figure~\ref{fig:G2} shows analogous plots for $\varepsilon$~\eqref{eq:error_norm} with respect to $\ntriangles$ for $d=2$ in~\eqref{eq:G_mms}.  These plots also indicate that the solution is second-order accurate.  Table~\ref{tab:G2} shows the maximum rank, $\max\rank$, of~\eqref{eq:zjv_mms} for each of the simulations, across the different mesh sizes.  These values are also low, albeit slightly higher than the values in Table~\ref{tab:G1}.  The higher values in Table~\ref{tab:G2} are due to the greater relative weight of the matrix diagonal that arises from this choice of $G_\text{MS}$.  Higher values of $d$ would be expected to increase $\rank$, but $\rank$ would most likely remain small relative to $\nbasis$.  Additionally, higher values of $d$ require more quadrature points for exact integration, increasing the computational cost and potentially requiring finer meshes for the discretizations to be in the asymptotic region.

\reviewerTwo{Figures~\ref{fig:G1} and~\ref{fig:G2} show that the uniform mesh has a lower error for $\mathbf{Z}^\Phi$, whereas the twisted mesh has a lower error for $\mathbf{Z}_\text{MS}$ and $\mathbf{Z}^\mathbf{A}$.  These performance differences are due to fortuitous mesh concentrations in regions with higher gradients.  Simulations using the uniform mesh enter the asymptotic region faster, due to the regularity of the mesh.}

\begin{figure}[!t]
\centering
\includegraphics[scale=.64,clip=true,trim=0in 0in 0.5in 0in]{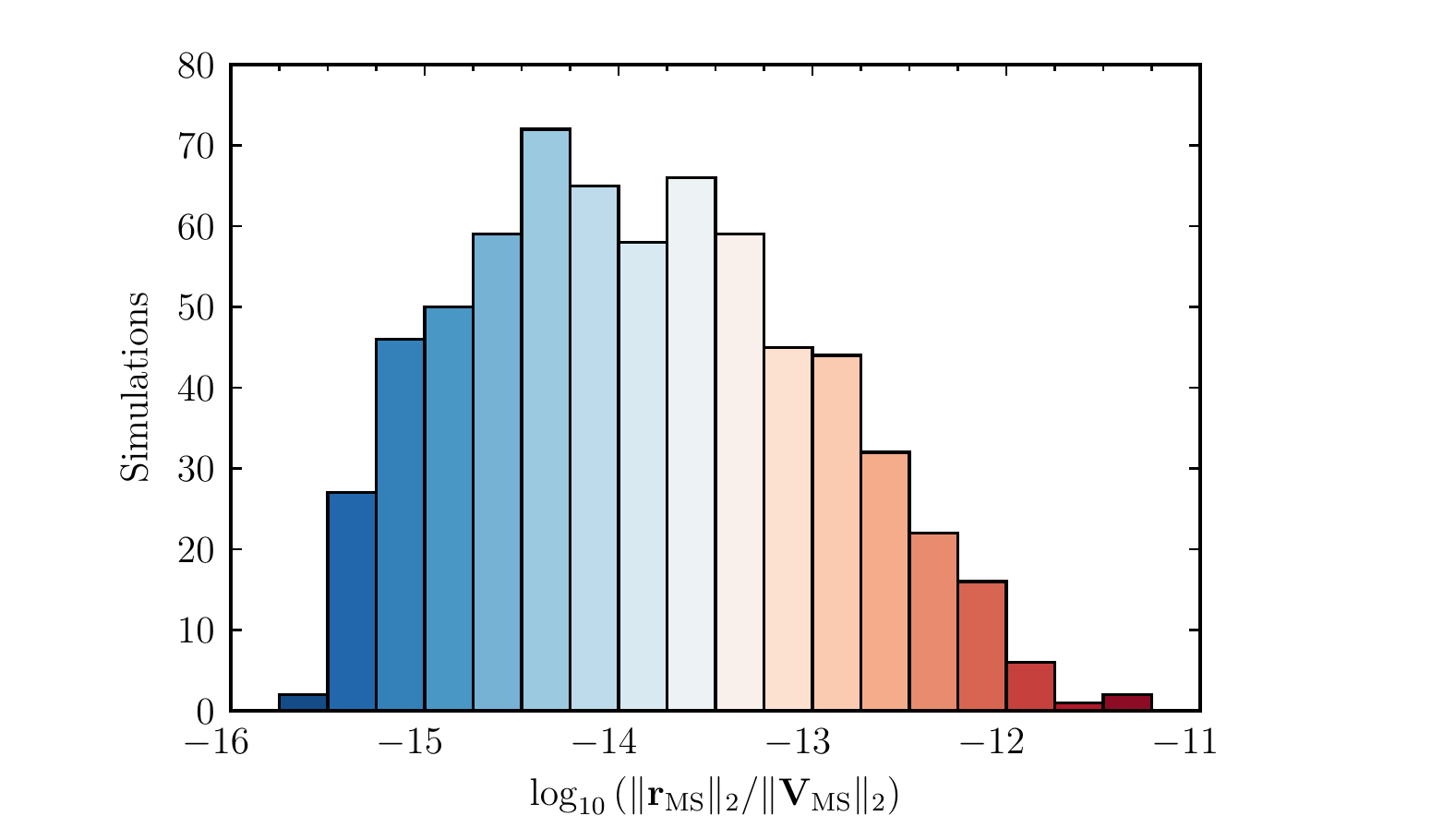}
\caption{$\mathbf{r}_\text{MS}$~\eqref{eq:zjv_mms} for the simulations in Figures~\ref{fig:G1} and~\ref{fig:G2}.}
\label{fig:res}
\end{figure}

Finally, Figure~\ref{fig:res} shows the distribution of the pooled residuals $\mathbf{r}_\text{MS}$~\eqref{eq:zjv_mms} from each of the simulations plotted in Figures~\ref{fig:G1} and~\ref{fig:G2}.  These low values confirm~\eqref{eq:zjv_mms} has been satisfied.

\begin{table}[!b]
\centering
\begin{tabular}{c c c c c c c}
\toprule
& \multicolumn{3}{c}{Uniform} & \multicolumn{3}{c}{Twisted} \\
 \cmidrule(lr){2-4} \cmidrule(lr){5-7}
$\theta$
& $\mathbf{Z}_\text{MS}$ & $\mathbf{Z}^\mathbf{A}$ & $\mathbf{Z}^\Phi$ 
& $\mathbf{Z}_\text{MS}$ & $\mathbf{Z}^\mathbf{A}$ & $\mathbf{Z}^\Phi$
\\
\midrule
$\pz\pz0^\circ$ & \pz8           & \pz8           & \pz2 & \pz8           & \pz8           & \pz2 \\
  $\pz45^\circ$ & \rereading{13} & \rereading{13} & \pz3 & \rereading{13} & \rereading{13} & \pz3 \\
  $\pz90^\circ$ & \rereading{13} & \rereading{13} & \pz3 & \rereading{13} & \rereading{13} & \pz3 \\
    $135^\circ$ & \rereading{13} & \rereading{13} & \pz3 & \rereading{13} & \rereading{13} & \pz3 \\
\bottomrule
\end{tabular}
\caption{Maximum rank of~\eqref{eq:zjv_mms}, $\max\rank$, across meshes for $d=1$ in $G_\text{MS}$~\eqref{eq:G_mms}.}
\vskip-\dp\strutbox
\label{tab:G1}
\end{table}

\begin{table}[!b]
\centering
\begin{tabular}{c c c c c c c}
\toprule
& \multicolumn{3}{c}{Uniform} & \multicolumn{3}{c}{Twisted} \\
 \cmidrule(lr){2-4} \cmidrule(lr){5-7}
$\theta$
& $\mathbf{Z}_\text{MS}$ & $\mathbf{Z}^\mathbf{A}$ & $\mathbf{Z}^\Phi$ 
& $\mathbf{Z}_\text{MS}$ & $\mathbf{Z}^\mathbf{A}$ & $\mathbf{Z}^\Phi$
\\
\midrule
$\pz\pz0^\circ$ & 18 & 18 & \pz7 & 18 & 18 & \pz7 \\
  $\pz45^\circ$ & \rereading{31} & \rereading{31} &   11 & \rereading{31} & \rereading{31} &   11 \\
  $\pz90^\circ$ & \rereading{31} & \rereading{31} &   11 & \rereading{31} & \rereading{31} &   11 \\
    $135^\circ$ & \rereading{31} & \rereading{31} &   11 & \rereading{31} & \rereading{31} &   11 \\
\bottomrule
\end{tabular}
\caption{Maximum rank of~\eqref{eq:zjv_mms}, $\max\rank$, across meshes for $d=2$ in $G_\text{MS}$~\eqref{eq:G_mms}.}
\vskip-\dp\strutbox
\label{tab:G2}
\end{table}

\subsection{Incorrect Implementation} 

In addition to showing that the approach of Section~\ref{sec:mms} confirms the expected accuracy in the absence of coding errors, it is \reviewerTwo{important} to show that the approach can detect coding errors as well.  Given the low ranks of the constraints in Tables~\ref{tab:G1} and~\ref{tab:G2}, it is important to ensure the admitted solution is not correct and does not achieve the expected order of accuracy in the presence of coding errors.

We consider four coding errors to test the approach:
\begin{enumerate}[leftmargin=*,labelindent=\parindent,label=Case~\arabic*:,ref=\arabic*]
\item \label{ce:k} \textit{Incorrect value of $k$}.  $k$ is increased by 1\%, incorrectly weighting the contribution to $\mathbf{Z}_\text{MS}$ from $\mathbf{Z}_\mathbf{A}$.
\item \label{ce:quad} \textit{Incorrect quadrature weights}.  The weights are increased by 1\%, making the solutions to the integrals inconsistent.
\item \label{ce:mat_elem} \textit{Incorrect matrix entry}.  $Z_{\text{MS}_{1,2}}$ is increased by 1\%.
\item \label{ce:area} \textit{Incorrect triangle areas}.  Instead of using the correct areas for the triangle, the areas associated with the uniform mesh are used in the basis function evaluations.
\end{enumerate}

\begin{figure}[!t]
\centering
\includegraphics[scale=.64,clip=true,trim=2.3in 0in 2.8in 0in]{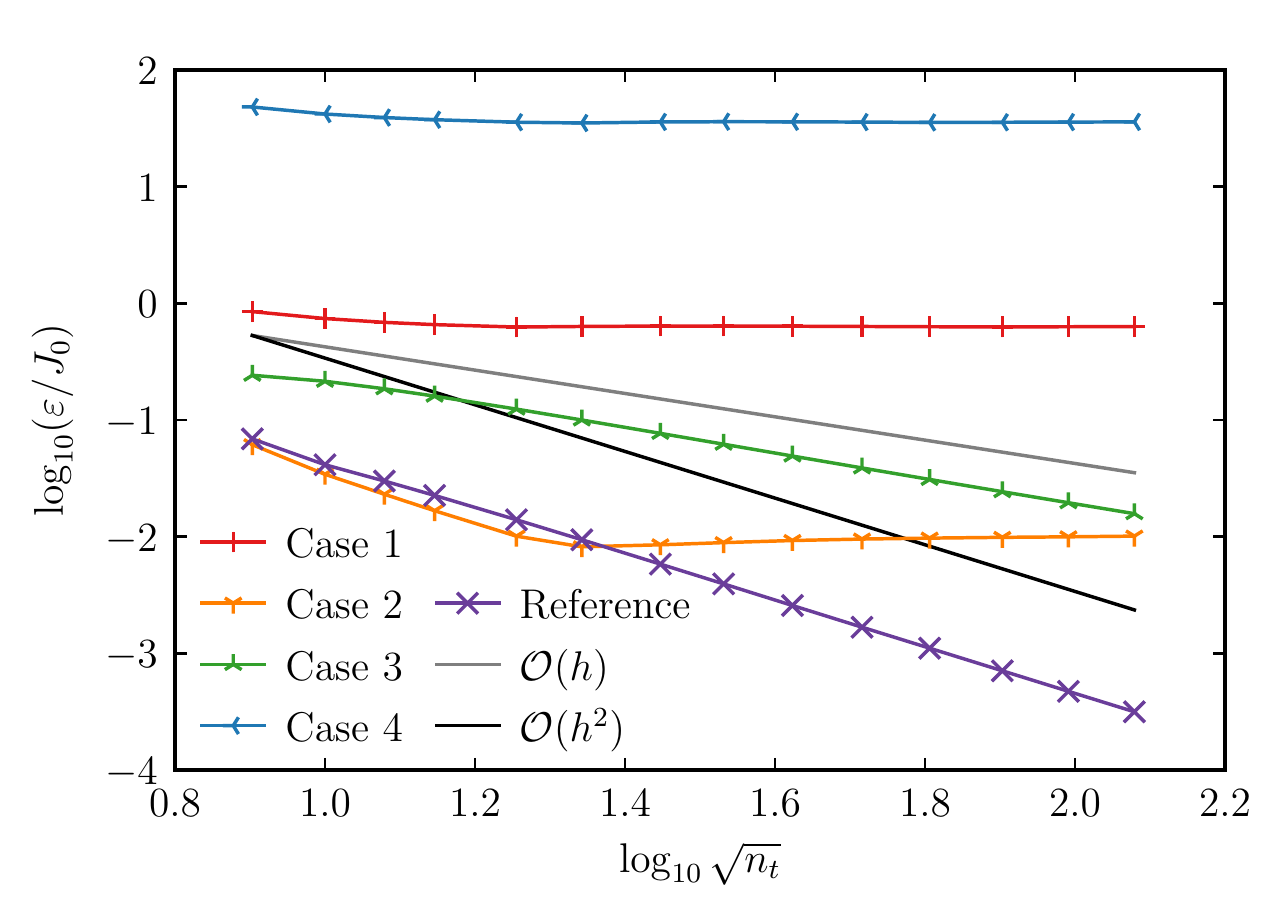}
\caption{Error $\varepsilon$~\eqref{eq:error_norm} for multiple coding errors.}
\label{fig:bugs_linf}
\end{figure}

For the twisted mesh, with both contributions to $\mathbf{Z}_\text{MS}$, $d=1$ in $G_\text{MS}$~\eqref{eq:G_mms}, and $\theta=45^\circ$, Figure~\ref{fig:bugs_linf} shows the errors of the four coding errors, as well as the correct result from Figure~\ref{fig:G1:45}.  Cases~\ref{ce:k}, \ref{ce:quad}, and \ref{ce:area} compromise the consistency of the governing equations, such that the error does not decrease with mesh refinement.  On the other hand, Case~\ref{ce:mat_elem} remains consistent but increases the error and slows its rate of decay to first order.  \reviewerTwo{The error for this case is able to decay because, as the mesh is refined, the spatial location of this fixed-column-index error moves toward the corner of the domain, where the normal components of the surface current vanish.  Additional details are provided in~\ref{sec:app}.  Nonetheless, by decaying at a slower rate than expected, this coding error is detected.}

\section{Conclusions}
\label{sec:conclusions}

As outlined in the introduction, there are several combinations through which the different error sources in the MoM implementation of the EFIE can interact.  \reviewerTwo{Combinations that avoid the solution-discretization error are not verifiable, and combinations that verify the numerical-integration error can be impractical due to the difficulty of computing accurate reference solutions for singular integrals}.  In this paper, we presented an approach through which the order of the solution-discretization error, Error Source~\ref{err:sol_disc}, can be verified.  Through this approach, we manufactured both the surface current and Green's function, and reduced the arising singular system of equations to a set of constraints for an optimization problem that selects the permissible solution closest to the manufactured solution.  We demonstrated the effectiveness of this approach for properly and improperly coded examples.

\reviewerOne{Though we demonstrated this approach using a single basis-function choice, this approach can be extended to include other basis-function choices with minimal implications.} This approach can additionally be extended to include the domain-discretization error, Error Source~\ref{err:dom_disc}, to make Combination 4, \reviewerTwo{which accounts for the domain-discretization and solution-discretization errors,} verifiable.  Therefore, verifying \reviewerTwo{combinations that account for the numerical-integration error} remains an open challenge for practical problems, requiring an acceptable balance of integration accuracy and computational cost.
\section*{Acknowledgments} 
\label{sec:acknowledgments}
The authors thank Justin Owen for his insightful feedback.
This paper describes objective technical results and analysis. Any subjective views or opinions that might be expressed in the paper do not necessarily represent the views of the U.S. Department of Energy or the United States Government.
Sandia National Laboratories is a multimission laboratory managed and operated by National Technology and Engineering Solutions of Sandia, LLC, a wholly owned subsidiary of Honeywell International, Inc., for the U.S. Department of Energy's National Nuclear Security Administration under contract DE-NA-0003525.

\appendix
\renewcommand{\thesection}{Appendix~\Alph{section}}
\section{Implication of Matrix-Element Coding Error}
\label{sec:app}

The purpose of this appendix is to illustrate the implications of a coding error limited to a matrix element.
From~\eqref{eq:zjv_mms}, the original system of equations is
\begin{align}
\mathbf{Z}_\text{MS} \mathbf{J}^h = \mathbf{V}_\text{MS}.
\label{eq:correct}
\end{align}
Introducing a coding error $\delta$ in element $(i,j)$ of $\mathbf{Z}_\text{MS}$, the erroneous matrix is
\begin{align}
\tilde{\mathbf{Z}}_\text{MS} = \mathbf{Z}_\text{MS} + \delta \mathbf{e}_i\mathbf{e}_j^T,
\label{eq:incorrect_Z}
\end{align}
where $\mathbf{e}_i$ denotes the standard unit vector with zero-valued elements, except at element $i$, where the value is unity.
Inserting~\eqref{eq:incorrect_Z} into~\eqref{eq:correct} yields the modified equation
\begin{align}
\tilde{\mathbf{Z}}_\text{MS} \tilde{\mathbf{J}}^h = \mathbf{V}_\text{MS}.
\label{eq:incorrect}
\end{align}

From~\eqref{eq:error}, the error in~\eqref{eq:correct} is
\begin{align}
\|\mathbf{J}^h - \mathbf{J}_n \|\le Ch^p,
\label{eq:correct_error}
\end{align}
whereas the error in~\eqref{eq:incorrect} is
\begin{align}
\|\tilde{\mathbf{J}}^h - \mathbf{J}_n\| \le \tilde{C}h^{\tilde{p}}.
\label{eq:incorrect_error}
\end{align}

Using the Sherman--Morrison formula,
\begin{align*}
\tilde{\mathbf{J}}^h 
= \tilde{\mathbf{Z}}_\text{MS}^{-1} \mathbf{V}_\text{MS} 
= \left(\mathbf{Z}_\text{MS}^{-1} - \frac{\delta\mathbf{Z}_\text{MS}^{-1}\mathbf{e}_i\mathbf{e}_j^T\mathbf{Z}_\text{MS}^{-1}}{1+\delta\mathbf{e}_j^T\mathbf{Z}_\text{MS}^{-1}\mathbf{e}_i}\right)\mathbf{V}_\text{MS}
= \mathbf{J}^h - \frac{\delta\mathbf{Z}_\text{MS}^{-1}\mathbf{e}_i\mathbf{e}_j^T \mathbf{J}^h}{1+\delta\mathbf{e}_j^T\mathbf{Z}_\text{MS}^{-1}\mathbf{e}_i},
\end{align*}
such that
\begin{align}
\|\tilde{\mathbf{J}}^h - \mathbf{J}^h\|
\le
\frac{\|\delta\mathbf{Z}_\text{MS}^{-1}\mathbf{e}_i\||\mathbf{e}_j^T \mathbf{J}^h|}{|1+\delta\mathbf{e}_j^T\mathbf{Z}_\text{MS}^{-1}\mathbf{e}_i|}.
\label{eq:rate1}
\end{align}
In~\eqref{eq:rate1}, $\|\delta\mathbf{Z}_\text{MS}^{-1}\mathbf{e}_i\|\approx C_1 h^q$ and $|\delta\mathbf{e}_j^T\mathbf{Z}_\text{MS}^{-1}\mathbf{e}_i|\approx C_2 h^q$, where $q$ is negative.  If $|\mathbf{e}_j^T \mathbf{J}^h|\le C_3 h^r$, as $h\to 0$,
\begin{align*}
\|\tilde{\mathbf{J}}^h - \mathbf{J}^h\|
\le
C' h^r,
\end{align*}
such that
\begin{align*}
\|\tilde{\mathbf{J}}^h - \mathbf{J}_n \|
\le 
\|\tilde{\mathbf{J}}^h - \mathbf{J}^h\| + \|\mathbf{J}^h - \mathbf{J}_n\|
\le
C' h^r + Ch^p
\le
\tilde{C} h^{\min\{r,p\}}.
\end{align*}
Therefore, $\tilde{p}=\min\{r,p\}$ in~\eqref{eq:incorrect_error}.

For a fixed index $j$, as the mesh is refined, the spatial location moves toward the corner of the domain where the normal components of the surface current vanish.  Therefore, $r$ is positive. 
In Case~\ref{ce:mat_elem}, the choice of $\mathbf{J}_\text{MS}$ results in $r=1$, due to the linear arguments of the sinusoidal functions; therefore, $\tilde{p}=1$.

\addcontentsline{toc}{section}{\refname}
\bibliographystyle{elsarticle-num}
\bibliography{quadrature.bib}

\end{document}